\documentclass[a4paper,11pt]{article}
\pdfoutput=1 % if your are submitting a pdflatex (i.e. if you have
\usepackage{jinstpub} % for details on the use of the package,
\usepackage[utf8]{inputenc} % usually not needed (loaded by default)
\usepackage[T1]{fontenc}
\usepackage{natbib,hyperref}
\usepackage{lineno}
\usepackage{subcaption}
\usepackage{graphicx}
\usepackage{textcomp}
\usepackage{mathtools}
\usepackage{siunitx}
\DeclarePairedDelimiterXPP\BigOSI[2]%
  {\mathcal{O}}{(}{)}{}%
  {\SI{#1}{#2}}

\title{\boldmath First ClearMind gamma detector prototype for TOF-PET imaging}

\author[a,1]{A. Galindo-Tellez \note{Corresponding author}}
\author[a,b]{, V. Sharyy}
\author[a,2]{, C.-H. Sung \note{Currently at GE}}
\author[a]{, M. Follin}
\author[c]{, L. Cappellugola}
\author[c,3]{, S. Curtoni  \note{Currently at Alten}} 
\author[c]{, M. Dupont}
\author[c]{, C. Morel}
\author[d]{, D. Breton}
\author[d]{, J. Maalmi}
\author[a,b]{and D. Yvon}
\affiliation[a]{Universit\'e Paris-Saclay, CEA, IRFU, D\'epartement de Physique des Particules, Gif-sur-Yvette, France}
\affiliation[b]{Universit\'e Paris-Saclay, CEA, CNRS, INSERM, Laboratoire d’Imagerie Biom\'edicale Multimodale Paris Saclay, Orsay, France}
\affiliation[c]{Aix-Marseille Université, CNRS/IN2P3, CPPM, Marseille, France}{}
\affiliation[d]{Université Paris-Saclay, CNRS, IJCLab, Orsay, France}{}
\emailAdd{aline.galindotellez@cea.fr}

\abstract{The ClearMind project aims to develop a TOF-PET position-sensitive detection module optimized for time and spatial resolutions and detection efficiency. For this, we use a 59 mm $\times$ 59 mm $\times$ 5 mm monolithic PbWO$_4$ (PWO) crystal, which is encapsulated within a commercial Micro-Channel Plate Photomultiplier tube MAPMT253 with a bialkali photocathode directly deposited on the crystal. We report the proof of concept of the directly deposited of a bialkali photocathode on a PWO crystal and its stability over time. The full calibration of the ClearMind photodetector module in the single-photoelectron regime is described. We measured a time resolution of 70 ps FWHM using a 20 ps pulsed laser. We present the performance of the prototype used in coincidence with a 3 $\times$ 3 $\times$ 3 mm$^3$ LYSO:Ca,Ce crystal readout by a SiPM. We obtained a coincidence time resolution of 350 ps FWHM, a spatial resolution of 4 to 5 mm, and a detection efficiency of 28 \%, consistent with Monte Carlo simulations of the ClearMind detector module.
}

\arxivnumber{2312.13169} % only if you have one

\begin{document}
\maketitle
\flushbottom

\section{Introduction}
\label{sec:intro}
Positron emission tomography (PET) is a powerful imaging tool used in nuclear medicine to visualize and measure metabolic activity at the molecular level \cite{Jiang2019SensorsFP}. The PET image is obtained by injecting a positron-emitting radiopharmaceutical into the patient's bloodstream. The tracer will bind specifically to the consuming organ and decay by emitting a positron that annihilates with a nearby electron, resulting in the antiparallel emission of two 511 keV annihilation photons detected in coincidence. 
A line of response (LOR) is defined by connecting the two detection points of the coincidence. The acquisition of millions of coincidences allows us to reconstruct the 3D activity concentration of the radiopharmaceutical. Additionally, in the time-of-flight (TOF) technique, the measurement of the difference in detection time of the two annihilation photons helps to pre-localize the annihilation position to a smaller region along the corresponding LOR, thus improving the signal-to-noise ratio of the final image \cite{Ullah2016InstrumentationFT, Snyder1981,Budinger1983}.

The coincidence time resolution (CTR) quantifies the FWHM of the time difference distribution in the TOF measurements, which is proportional to the uncertainty on the localization of the annihilation position along the LORs.~\cite{8049484}. Among the benefits of CTR improvement, we can highlight the possibility of reducing the radiopharmaceutical activity (i.e., the dose administered to the patient) or the reduction of the scan duration, improving patient throughput \cite{Lecoq2020}.
 
Best current commercial PET scanners achieve CTR values of ~215 ps \cite{vanSluis2019PerformanceCO}.

The ClearMind project targets 511-keV gamma detection with improved timing performances (CTRs down to tens of picoseconds), without compromising the other detector performances, i.e. spatial resolutions down to a few cubic millimeters and good detection efficiency. We designed a detector with a large active surface, a low-power readout electronic compact enough to be conveniently integrated into a clinical PET design, and a technology that shows motivating performances. Thanks to its design, the ClearMind prototype (CMP) is a promising candidate to improve the CTR of PET detectors \cite{Yvon_2020,Follin_2021}.

In this work, we describe the first detector developed by the ClearMind project, the CMP. In Section 2, we describe the design of the CMP along with its readout system. 
In Section 3, we study one of the main components of our detector design, a photocathode directly deposited on a PWO crystal. A test cell is used to measure photocathode efficiency and determine its stability versus time. The same test bench is used to measure the CMP photon detection efficiency. In Section 4, we detail the CMP photodetection performance in the single photoelectron regime, with an emphasis on gain, spatial, and time resolutions. In Section 5 we present our Monte Carlo model of the detector, built on the measured properties of detector components. We introduce the measurement setup of the CMP detection properties for 511-keV photons. We report measurements of efficiency, spatial, and time resolutions using a $^{22}$Na source.  Finally, in section 6 we discuss Monte Carlo predictions to detector measured properties. We present a deeper understanding of the detector, and how we plan to improve its properties with subsequent versions.

\section{Description of the CMP}
\label{sec:detector}
The CMP consists of a 59 mm $\times$ 59 mm $\times$ 5 mm PbWO$_4$ (PWO) crystal manufactured by CRYTUR \cite{crytur}, that is used as the entrance window of a standard Microchannel Plate  Photomultiplier Tube (MCP-PMT) MAPMT253 device from Photek Ltd. \cite{photek_MAPMT253}. 
%%%%%%
\begin{figure}[ht!]
    \begin{minipage}[b]{0.48\textwidth}
        \includegraphics[width=1\textwidth]{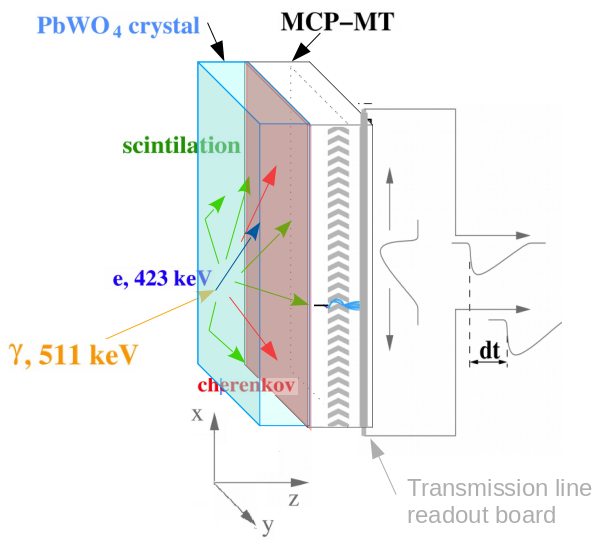}
        \caption{\label{fig:schematic}
        Diagram of the ClearMind detector module illustrating the detection process when a gamma-ray interacts via the photoelectric effect.}
    \end{minipage}
\hfill
    \begin{minipage}[b]{0.48\textwidth}
    \begin{center}
        \includegraphics[width=0.43\textwidth]{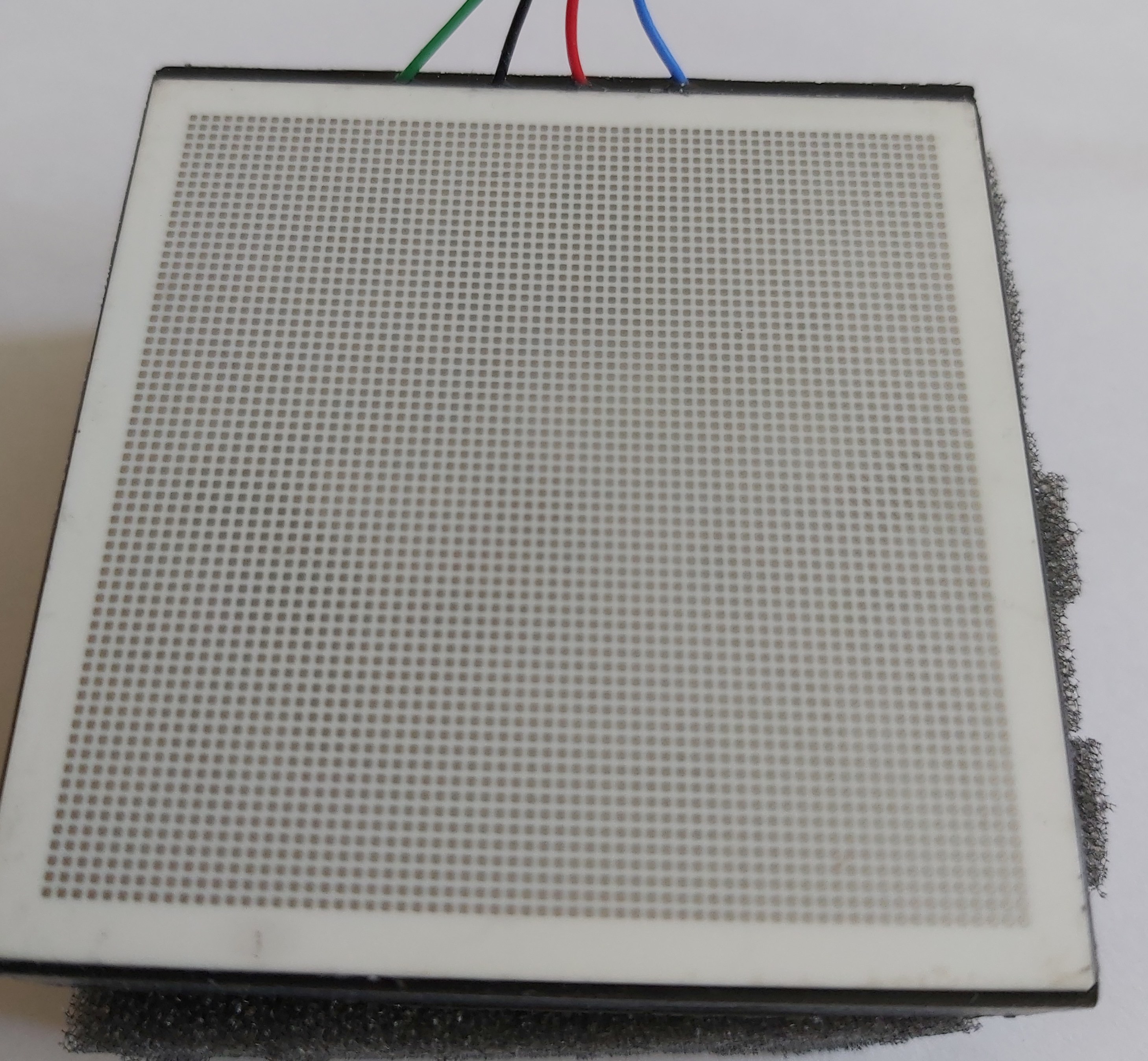}
        \includegraphics[width=0.53\textwidth]{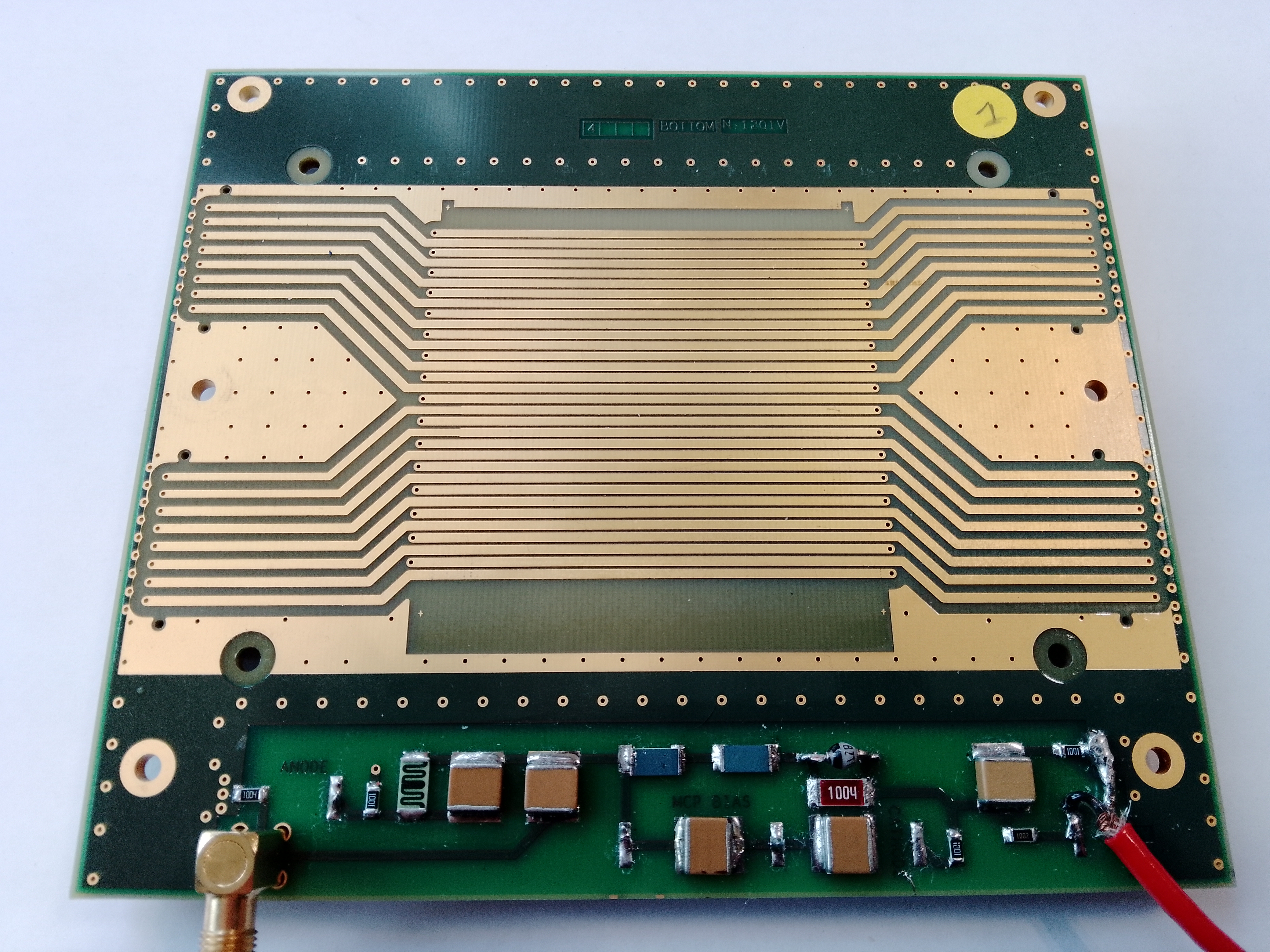}    
    \end{center}
    
    \caption{\label{fig:anodes}
        Photography of the anode matrix contact pads at the back of the CMP (\textbf{left}) and transmission lines of the readout PCB (\textbf{right}) used to interface anode matrix and readout amplifiers
    }
        
    \end{minipage}
\end{figure}
%%%%%%%
A passivation layer was deposited between the crystal and the bialkali photocathode to avoid chemical contamination of the photocathode.
 This PbWO$_4$ material has a high effective atomic number of 75.6 and a density of 8.28 g/cm$^3$. Hence it provides a short attenuation length of 9 mm for 511 keV gamma-rays and corresponds to an interaction probability of approximately 43\% within a 5 mm thick crystal \cite{XCOM}. When an interaction occurs by Compton or photoelectric effects (Fig.~\ref{fig:schematic}), the gamma-ray converts into an electron that is likely to generate both Cherenkov and scintillation photons. Among the optical photons that reach the photocathode, a fraction undergoes Fresnel reflections, while another fraction traverses the photocathode without being absorbed.  Those photons that are absorbed within the photocathode have a certain probability of being converted into photoelectrons.
Following the influence of the electric field, the photoelectrons undergo multiplication within the pores of the MCPs, inducing current pulses on the 4096 anode pads of the PMT. 
To minimize the number of electronic channels, we arranged the readout through 32 transmission lines (TLs) \cite{Follin_2022}. Each raw of  2 $\times$ 64 anodes pads is electrically connected to a 1.6~mm wide transmission line printed on the PCB through an anisotropic conductive rubber sheet, Inter-Connector\textregistered\ MT-type from Shin-Etsu \cite{interconnector}. 
The signals are read out and amplified at both ends of the transmission lines using dedicated two-stage amplifier boards (2 $\times$ 20 dB, 1 GHz, 50 Ohm) and recorded using a SAMPIC waveform digitizer \cite{BRETON2014, Delagnes2014, Delagnes:2015oda, Breton2016, Breton2020}. 

\section{Photon Detection Efficiency}
\label{sec:efficiency}
In order to be able to understand the behavior of our detector, we need to define two concepts related to the photon detection efficiency of the CMP.

The first is the Normal angle Photon Detection Efficiency (NPDE), which is the ratio between the number of photoelectrons extracted from the photocathode over the number of optical photons impinging at 90° on the optical window of the photodetector as a function of the wavelength. This concept is often referred to as the quantum efficiency in many photomultiplier technical data sheets.

The second is the Intrinsic photocathode Quantum Efficiency (IQE), defined as the ratio of the number of photoelectrons extracted from the photocathode over the number of optical photons impinging at the photocathode as a function of the wavelength.

A complete photocathode model, as presented in \cite{Motta2005a, Sung_2023}, requires parameterization of the optical index, absorption length, and extraction probability of photoelectrons produced within the photocathode as a function of wavelength. This is what has been implemented in the detector Monte Carlo simulation using data from \cite{Motta2005a}. Knowing the properties of the optical window, the photocathode refraction index, and when appropriate an optical passivation layer model, we are able to compute the Fresnel reflections at the optical interfaces. Then we compute the IQE of a photocathode at 90°, given an NPDE spectrum measured in \cite{Motta2005a}. Using published optical absorption lengths, and assuming a photocathode thickness of 25 nm, typical from a bi-alkali photocathode, we can compute the photoelectron extraction probability as a function of the wavelength. Variations in the NPDE, as measured in the ClearMind prototype, are implemented, by adjusting in the model a scaling factor on the photoelectron extraction probability.
The above computed photocathode properties thus allow us to simulate the optical behavior of the photodetector reliably. 

The ClearMind collaboration developed a test bench to measure the NPDE. The next section describes this test bench, the validation for the reference devices, and the NPDE measurement of the CMP.

\subsection{Quantum efficiency measurement setup}
The setup consists of a UV and VIS light source \cite{DH_2000} that goes through a monochromator \cite{monochromator} for wavelength selection. The output light enters inside a dark box via 1 mm$^{2}$ optical fiber and passes through a UV fused silica plano-convex lens (focal distance of 10 mm). The optical fiber output is placed at the focal point of the lens to produce a parallel light beam. An iris diaphragm is used to limit the beam size before the light reaches the detectors. The beam diameter, chosen to be 5 mm, fits within the sensitive area of the photodetector and reference photometers and was aligned to be located at the center of each photodetector. A computer-driven shutter allows the user to cut the optical beam when dark current measurements are needed.
%%%%%%%%
\begin{figure}[h!]
\centering
\includegraphics[width=0.8\textwidth]{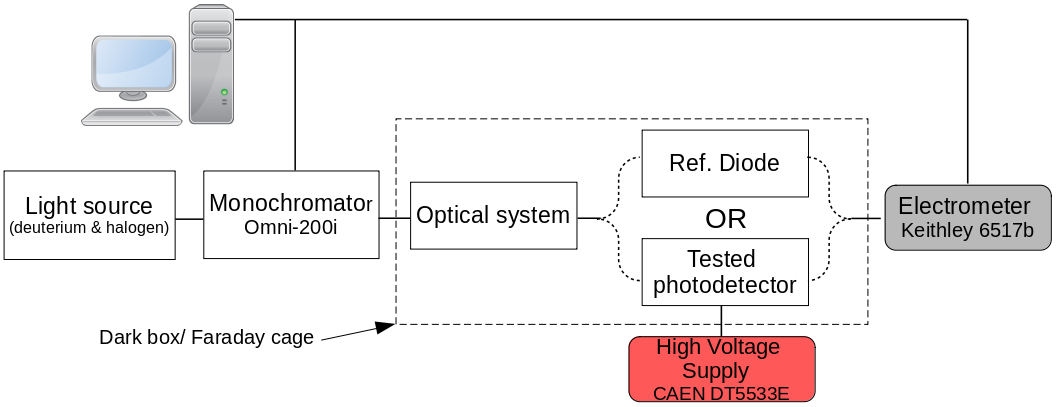}
\caption{Layout of the quantum efficiency experimental setup.}
\label{fig:QEschematic}
\end{figure}
%%%%%%%%%
The light source is warmed up for 20 min in order to optimize its stability. During the first measurement stage, the light spot illuminates the active surface of a reference diode. In the second measurement stage, the detector under test is illuminated. The currents are read out by an electrometer Keithley 6517b \cite{picoAmp} every second, five times per wavelength, and the average of these five measurements is stored on the PC.  The layout of the experimental setup for the NPDE measurement is shown in Fig. \ref{fig:QEschematic}.

%%%%%
\begin{figure}[h!]
\centering
\includegraphics[width=0.7\textwidth]{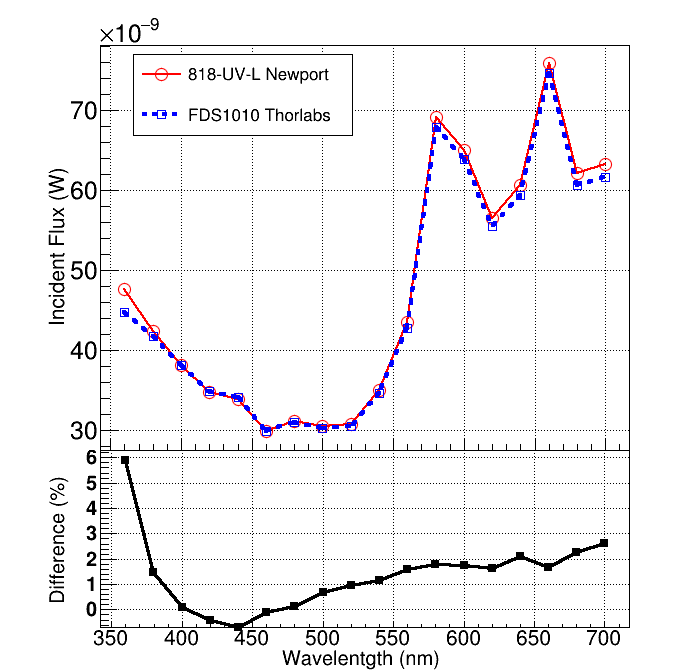}
\caption{\textbf{Top}: Incident flux measured by the reference photodiodes 818-UV/DB from Newport (continuous-red) and FS1010 from Thorlabs (dashed-blue) as a function of the wavelength. \textbf{Bottom}: Difference between the two curves expressed in \%.}
\label{fig:refDiodes}
\end{figure}
%%%%%%

\subsubsection{Validation of the reference devices}
To validate the reference devices, we utilized the testbench from Fig. \ref{fig:QEschematic}. The beam was characterized by an FS1010 Thorlabs reference diode \cite{ref_thorlabs} and an 818-UV/DB diode from Newport \cite{ref_newport} used under the same conditions. As the radiant sensitivity $S_k$ of a reference photodetector is given by the manufacturer, we can  compute the value of the incident radiant flux $L_p$ from the equation:
\begin{equation}
    L_p = \frac{I_k}{S_k}
\end{equation}
where $I_k$ is the measured photocurrent. The fluxes measured by both diodes are shown in Fig.~\ref{fig:refDiodes}. 
Both curves are in agreement within 2 \%, except for wavelengths below 380 nm, where a difference up to 6 \% is observed.  Our measurements agree with the suppliers' uncertainties of $\pm$ 5 \% for Thorlabs and less than 4 \% for Newport (1 \% at 350).
The consistency of both curves demonstrates the excellent behavior of the two diodes, the appropriate size and position of the light beam, and the light source stability.

\subsection{Test cell prototype}
\label{sec:testcell}
\begin{figure}[ht!]
\centering
\begin{subfigure}{0.48\textwidth}
    \includegraphics[width=\textwidth]{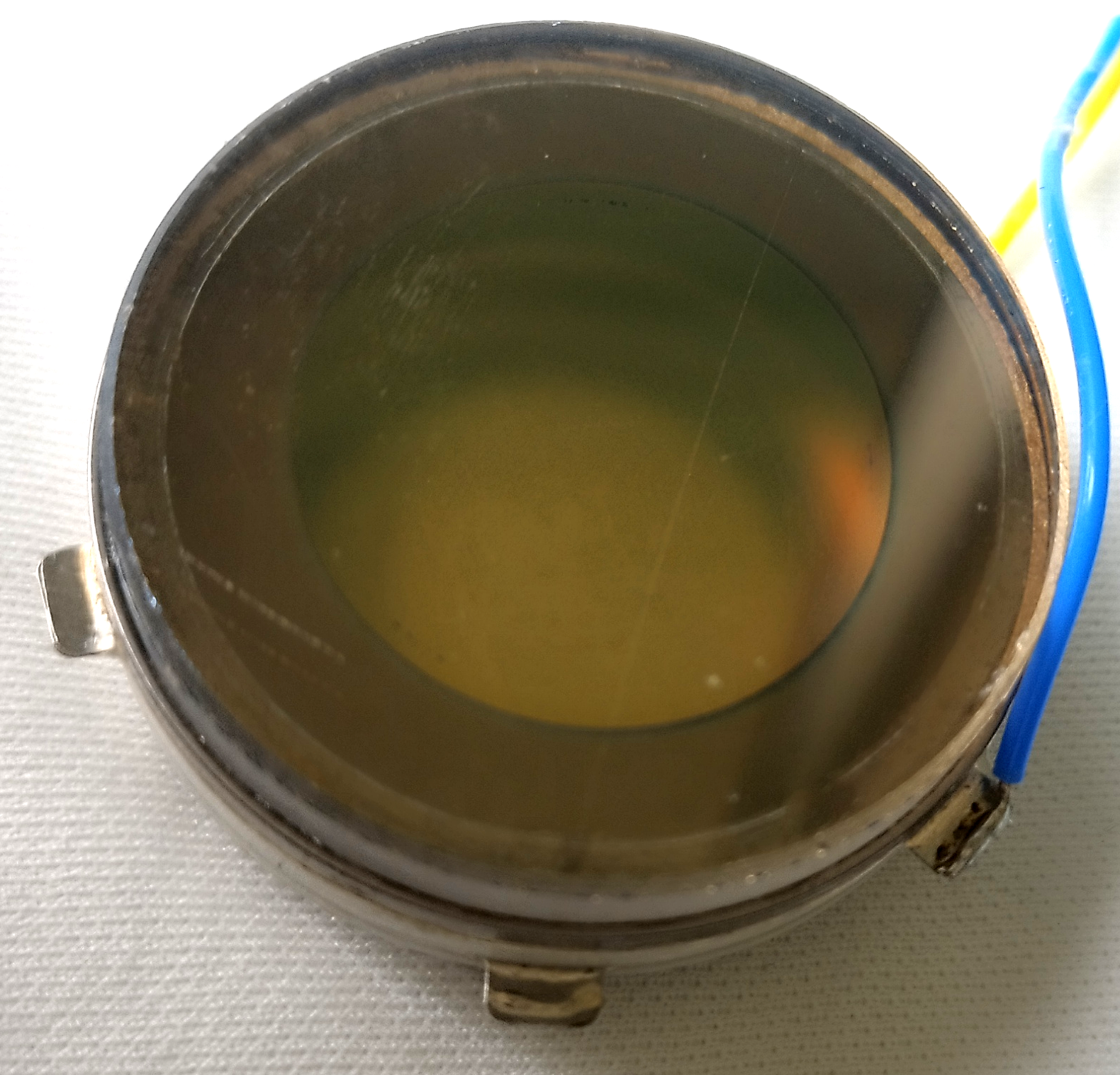}
\end{subfigure}
\hfill
\begin{subfigure}{0.48\textwidth}
    \includegraphics[width=\textwidth]{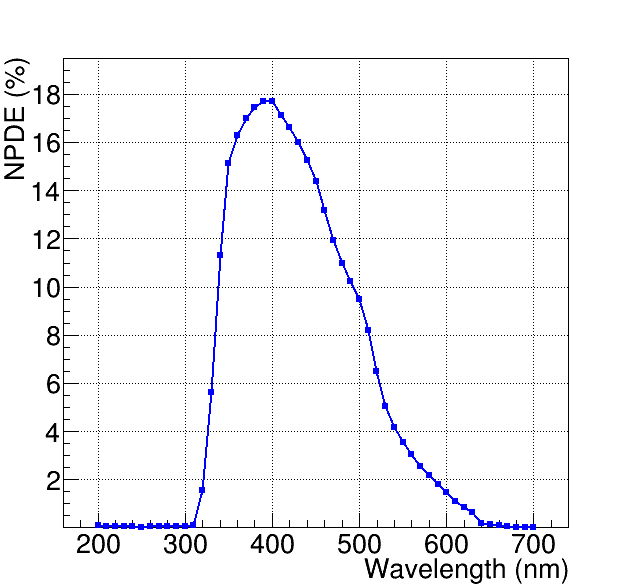}
\end{subfigure}
\caption{\textbf{Left : }PWO test cell. \textbf{Right : }Measured NPDE as a function of the wavelength from the PWO test cell.}
\label{fig:test_cell}
\end{figure}

Ahead of the fabrication of the CMP, test cells were built by Photek  Ltd. with a bialkali photocathode deposited on a passivated PWO optical window in order to test their performance and evaluate the feasibility of building a bigger and more complex prototype. A picture of a test cell is shown in Fig.\ref{fig:test_cell} left. 

%%%%%%

We tested devices with different passivation and photocathode configurations, as this is critical for the project. These technologies are covered by a non-diffusion agreement with Photek Ltd. and thus cannot be discussed in detail in this paper. The experimental setup is the same as the one shown in Fig. \ref{fig:QEschematic}.  The NPDE of the tested photodetector, $\eta$, is computed as:
%%%%%%%%%
\begin{equation}
    \eta(\lambda) = \eta_{ref} (\lambda) \frac{I_{test}(\lambda)}{I_{ref}(\lambda)},
    \label{eq:QE}
\end{equation}
%%%%%%%%%
where $\eta_{ref}$ is the quantum efficiency of the reference diode, $I_{test}$ is the current from the tested photodetector, $I_{ref}$ is the reference diode current, and $\lambda$ is the wavelength \cite{hamamatsu}. $I_{test}$ is measured directly by the test cell, and dark current is measured for each wavelength after measuring $I_{test}$ to subtract any current offset. We used the reference diode 818-UV/DB Newport to be sensitive below 350 nm. The current is directly measured at the photocathode's output, i.e. no multiplication is carried out, and the measurement drives directly to the NPDE determination.
The $\eta(\lambda)$ curve is shown in Fig. \ref{fig:test_cell} left. The NPDE peaks at 18 \% at a wavelength of 400 nm.
The observed cut for wavelengths below 350 nm is due to the PWO being no longer transparent. The cell which was built with similar technology as the CMP, was tested for the first time in 2021 and was found to be stable over 2 years.
Due to the light impinging at 90$^\circ$ with respect to the optical window, this measurement provides a convenient measurement of photocathode IQE, see Section. \ref{sec:efficiency}.

\subsection{CMP photon detection efficiency}
\label{sec:deteff}
The measurement of $I_{test}$ implies extracting the free electron produced at the photocathode. For the CMP, a voltage of $-200$~V  between the photocathode and the MCP input is applied. The light beam impinges perpendicularly to the center of the CMP optical window.  
%%%
\begin{figure}[h!]
\centering
\includegraphics[width=0.5\textwidth]{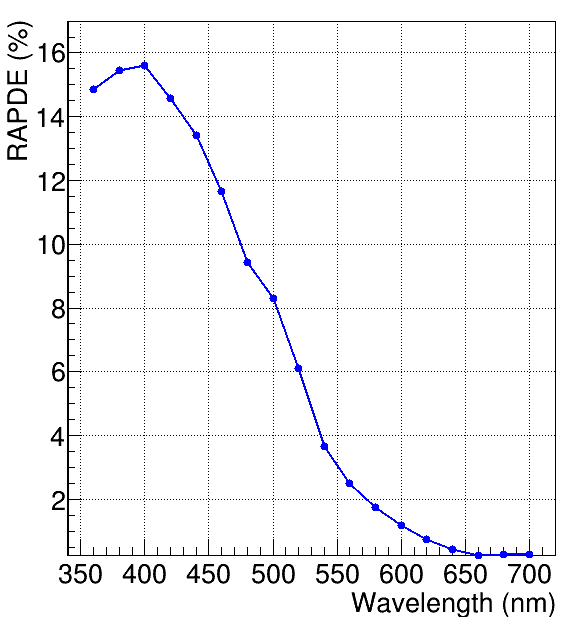}
\caption{Measured NPDE of the CMP. This measurement does not take into account the photoelectron collection efficiency at the MCP input, which is included afterward in our Monte Carlo simulation.}
\label{fig:QE_measured}
\end{figure}
%%%
As there is no need for charge multiplication, the MCP input and output are shorted-circuited, and the current is collected directly from it. The CMP anodes matrix is grounded to a transmission line board using a Shin-Etsu MT-type of Inter-Connector\textregistered~ in order to avoid floating anodes and spurious currents. The output current is low-pass filtered before being measured by the electrometer.
Fig. \ref{fig:QE_measured} shows the NPDE curve measured with the CMP, with a maximum value of 15.6 \% occurring at 400 nm. The measured value will be discussed in section \ref{sec:discusion}. 

\section{Photodetector calibration in the single photoelectron regime}
\label{sec:calibration}
\subsection{Test Setup}
\label{sec:calibration_setup}
For the calibration of the CMP, the readout is configured as described in Section \ref{sec:detector}. The total gain from the two amplification stages is 40 dB on each one of the 64 channels. The detector and the boards are placed inside a dark/EM-shielded box. A 20 ps pulsed laser PiLas by A. L. S. \cite{PiLas} is mounted on a 2D motion station, moved by two X-LRT0100AL-C linear stages from Zaber Technologies Inc. The laser beam, collimated with a 40 $\mu$m pin-hole, impinges normally to the CMP  crystal surface. The distance between the laser output and the pin-hole is $\sim$130 mm, and the distance between the pin-hole and the CMP window input is $\sim$10 mm. Given the diameter of the optical fiber, the light spot diameter on the optical window was calculated to be 50 $\mu$m. 

The data acquisition is triggered in coincidence between CMP pulses (with a threshold set to 1/4 of photoelectron) and the Pilas logic trigger output. With this configuration, the CMP detects a single photoelectron in only 2\% of the PiLas laser pulses. The random coincidence rate is negligible. Assuming a Poisson distribution for the number of photoelectrons detected, the ratio of two/one-photoelectron events amounts to 1 \%.

%%%%%%%
\begin{figure}[h!]
\centering
\begin{subfigure}{0.6\textwidth}
    \includegraphics[width=\textwidth]{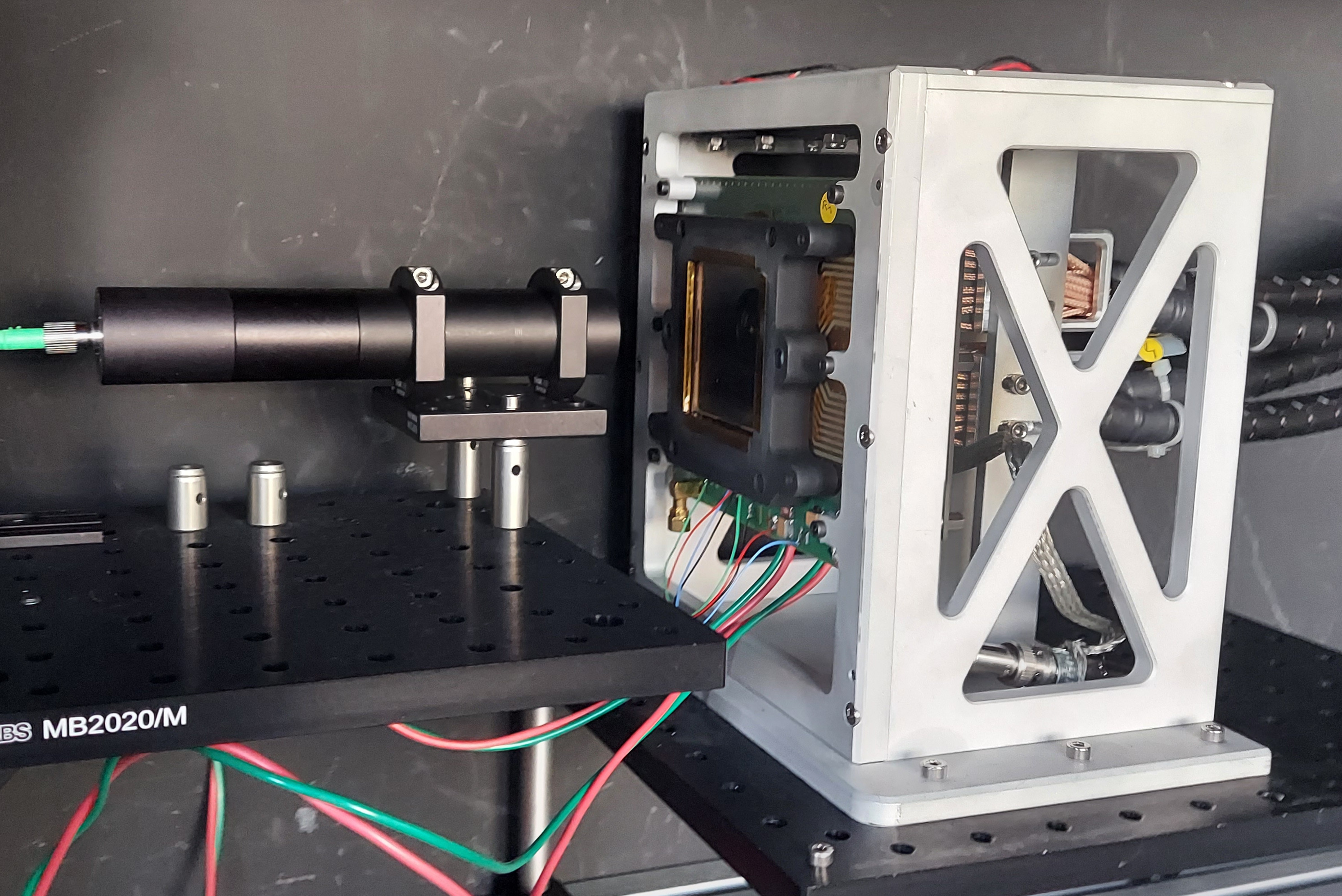}
\end{subfigure}
\hfill
\begin{subfigure}{0.39\textwidth}
    \includegraphics[width=\textwidth]{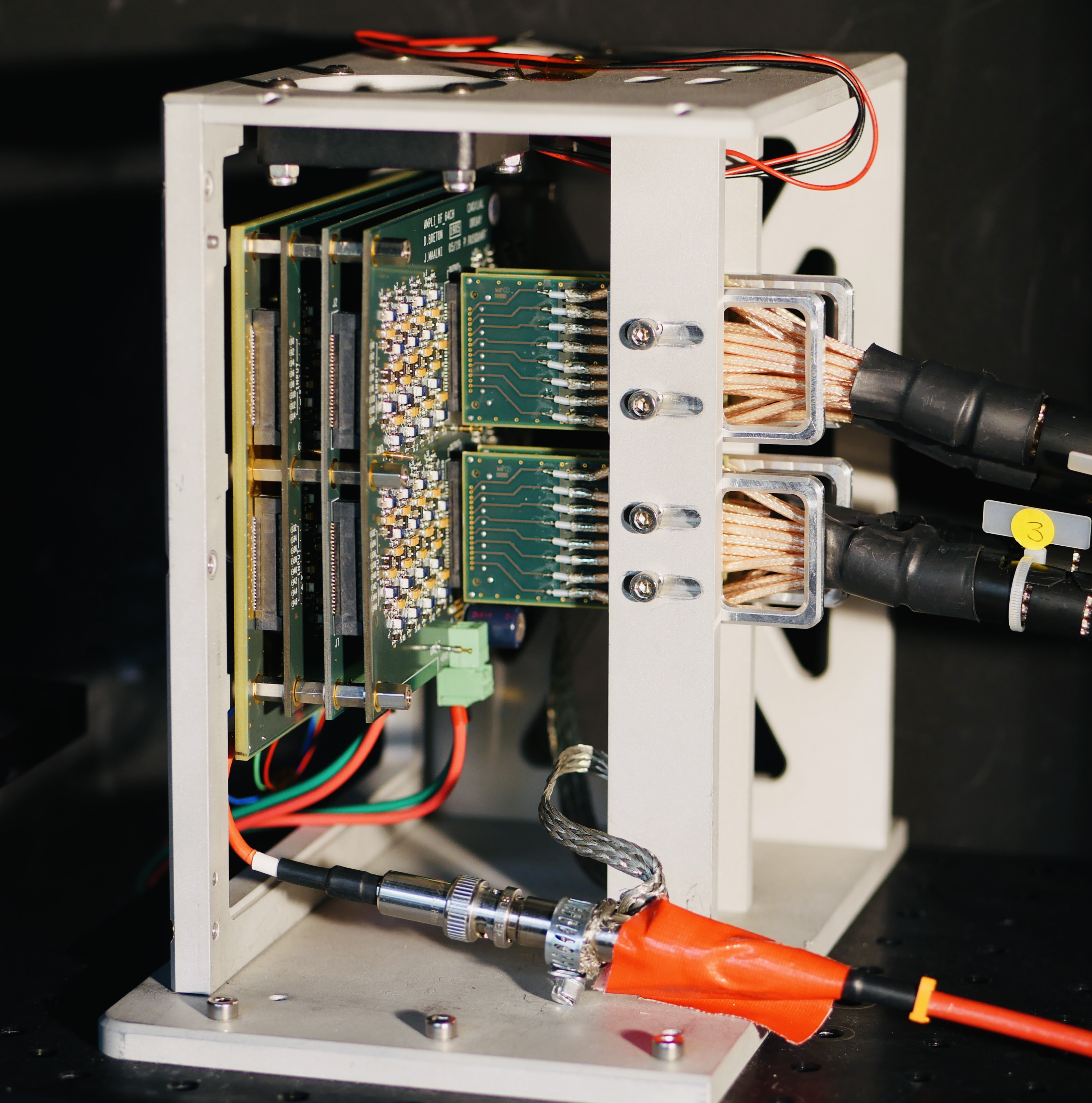}
\end{subfigure}       
\caption{Experimental setup for the CMP  calibration inside the dark box. \textbf{Left :} we can observe the laser mounted on the motor stage on the left side and the CMP on the right. \textbf{Right :} side view of the aluminum structure that holds the transmission line and amplification boards.}
\label{fig:setup}
\end{figure}
%%%%%%%

The signals were digitized by front-end boards from a SAMPIC crate @ 6.4 GS/s \cite{Delagnes:2015oda}  outside of the dark box and stored on a PC. We used a 50 mV threshold level for each channel, in order to reject noise. A picture of the experimental setup is shown in Fig. \ref{fig:setup}.

We performed a laser scan using 3 mm step sizes along the $x$-axis and 0.8 mm step sizes along the $y$-axis, which correspond to the parallel and perpendicular coordinates with respect to the TLs orientation. 
For appropriate sampling, the $y$-axis step size is chosen to be inferior to the pitch distance between the lines, which amounts to 1.66 mm. A total of 9000 different positions were scanned with an acquisition duration of 1 s for each. The events were acquired in coincidence with the output trigger of the laser. 
%%%
\begin{figure}[h!]
\centering
\includegraphics[width=0.55\textwidth]{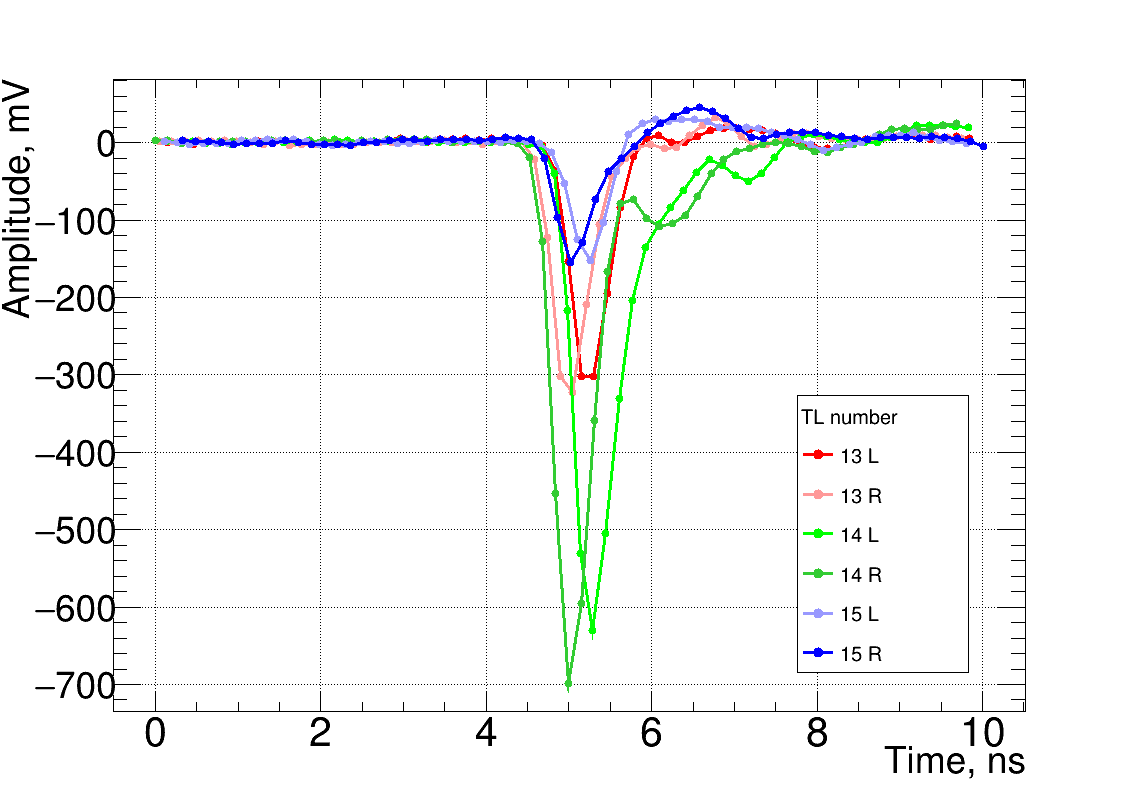}
\caption{Typical signals from the CMP, registered by the SAMPIC electronics.}
\label{fig:signal}
\end{figure}
%%%

%signal taken from file:///media/ag268352/Transcend/DATA/2022_09_13_laser/output_analysis_V2_advanceCalib/RESULTS/staging_position_419/gif/Coincidence_num_3.gif
Fig. \ref{fig:signal} shows some signals registered during the laser scan. Typical laser events trigger 1 to 3 lines, similar to the performances presented in \cite{Follin_2022}.

\subsection{Gain}
\begin{figure}[h!]
\centering
\includegraphics[width=0.50\textwidth]{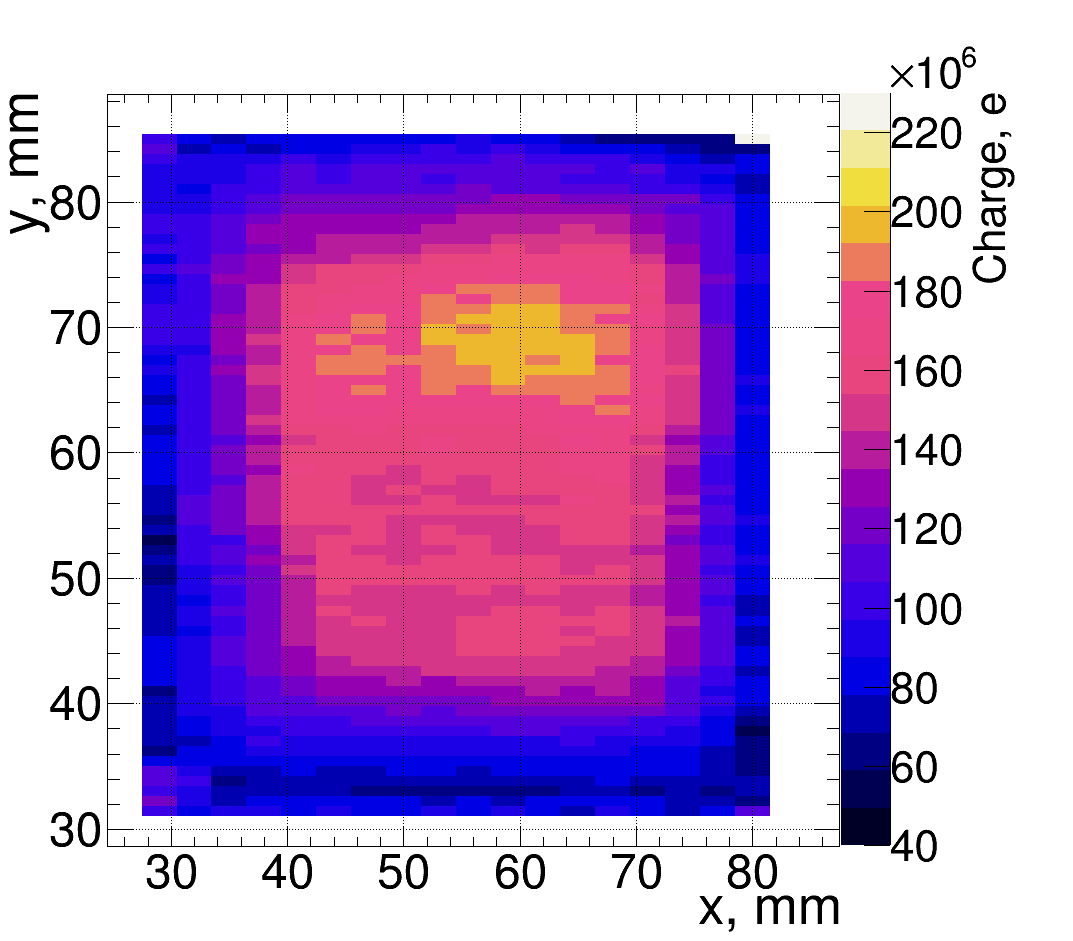}
\caption{Gain estimated across the CMP surface, after an amplification by a factor of 100. 
For this measurement, the two layers of MCP were biased under $-1620$~V. The higher gain values in the central part of the MCP result from an increase in the gap between the two MCP-PMT layers that impacts the number of channels of the second MCP layer involved in the amplification process. The high gain (yellow) area is not yet understood.}
\label{fig:charge}
\end{figure}

The MCP gain is computed at each scanned point by integrating the negative part of the pulses collected at the end of each triggered line and then summing them.
The gain depends on the MCP voltage bias and is chosen to have a 2 photoelectron pulse within the sensitivity range of the SAMPIC electronics. Fig. \ref{fig:charge} shows the number of electrons collected from the scanned CMP. The desirable pattern would be a uniform response within the surface of the PWO window. We observe a higher gain value in an upper half surface of 31 \% with respect to the gain averaged. We can also observe that the area near the boundaries has 50 \% less gain than the mean value. 
Overall, the full detection surface is responsive and active.
%%%%
%%%%

\subsection{Time resolution}
\label{sec:time_resolution}
\begin{figure}[h!]
\centering
\begin{subfigure}{0.50\textwidth}
    \includegraphics[width=\textwidth]{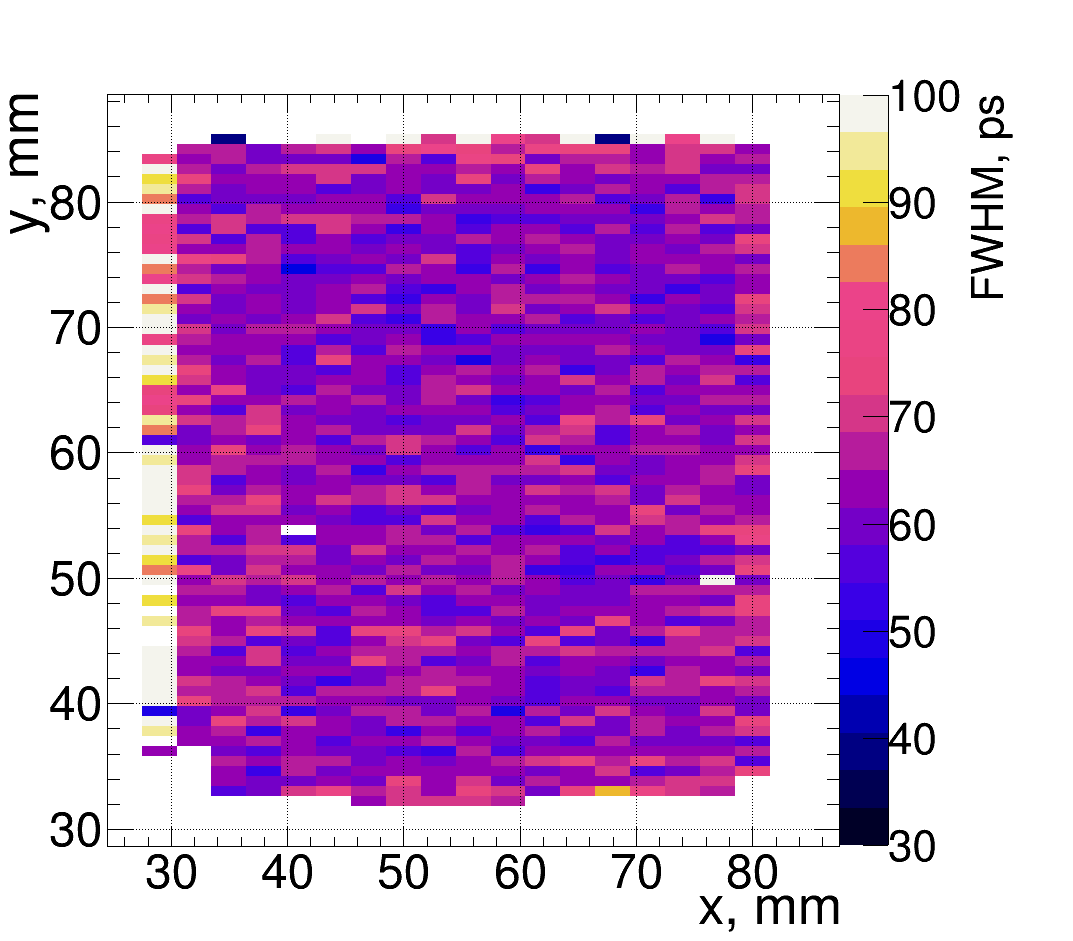}
\end{subfigure}
\hfill
\begin{subfigure}{0.46\textwidth}
    \includegraphics[width=\textwidth]{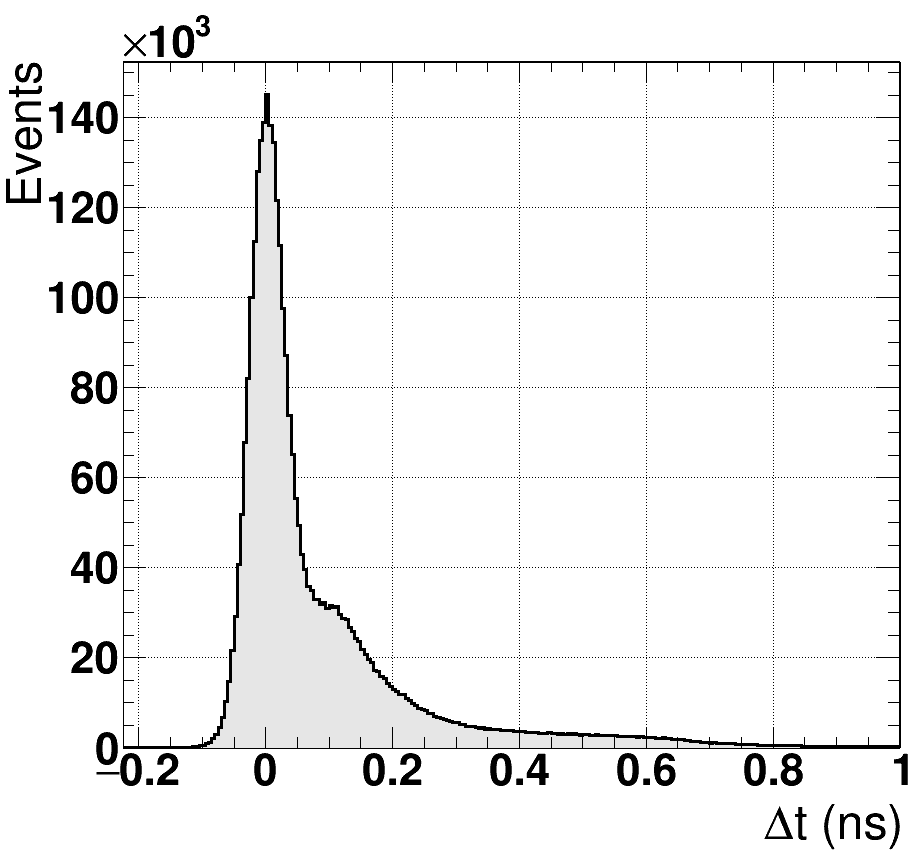}
\end{subfigure}
\caption{\textbf{Left : }FWHM of the time difference distribution between the CMP and the laser trigger, across the detection surface. \textbf{Right : }Time difference between the laser trigger signal and the CMP averaged over the detection surface. The main peak shows a width of 70 ps (FWHM).}
\label{fig:time_resolution}
\end{figure}

For each triggered line, a 50 \% constant fraction discrimination is used to determine the time of the registered pulses. The event time is selected as the timestamp of the line with the earliest recorded time. Detector time resolution is then computed by subtracting the laser trigger time. 
From every impinging position, the time resolution histogram is fitted using the sum of 3 Gaussian functions in order to adjust the time delays induced when the photoelectrons bounce once or twice before entering the MCP microchannels. Fig.\ref{fig:time_resolution} left, shows the FWHM of the first peak (Gaussian-shaped) of these distributions. A quite homogeneous distribution is observed along the sensitive surface of the CMP, with most bins ranging from 50 to 64 ps. The larger values happen at the edges of the sensitive surface.
%%%%%%%
 %%%%%%
Fig. \ref{fig:time_resolution} right, shows the time resolution distribution of the CMP averaged over the whole surface. The main peak has a width of 70 ps (FWHM) and includes 63 \% of the statistics. The events in the tail correspond to back-scattering electrons \cite{2018_Chen} that bounced on the first MCP surface. The time gap between the first Gaussian and the second one is directly related to the physical distance between the photocathode and the first MCP.

\subsection{Event position reconstruction}
\label{sec:positionReconstruction}
\begin{figure}[ht!]
\centering
\begin{subfigure}{0.45\textwidth}
    \includegraphics[width=\textwidth]{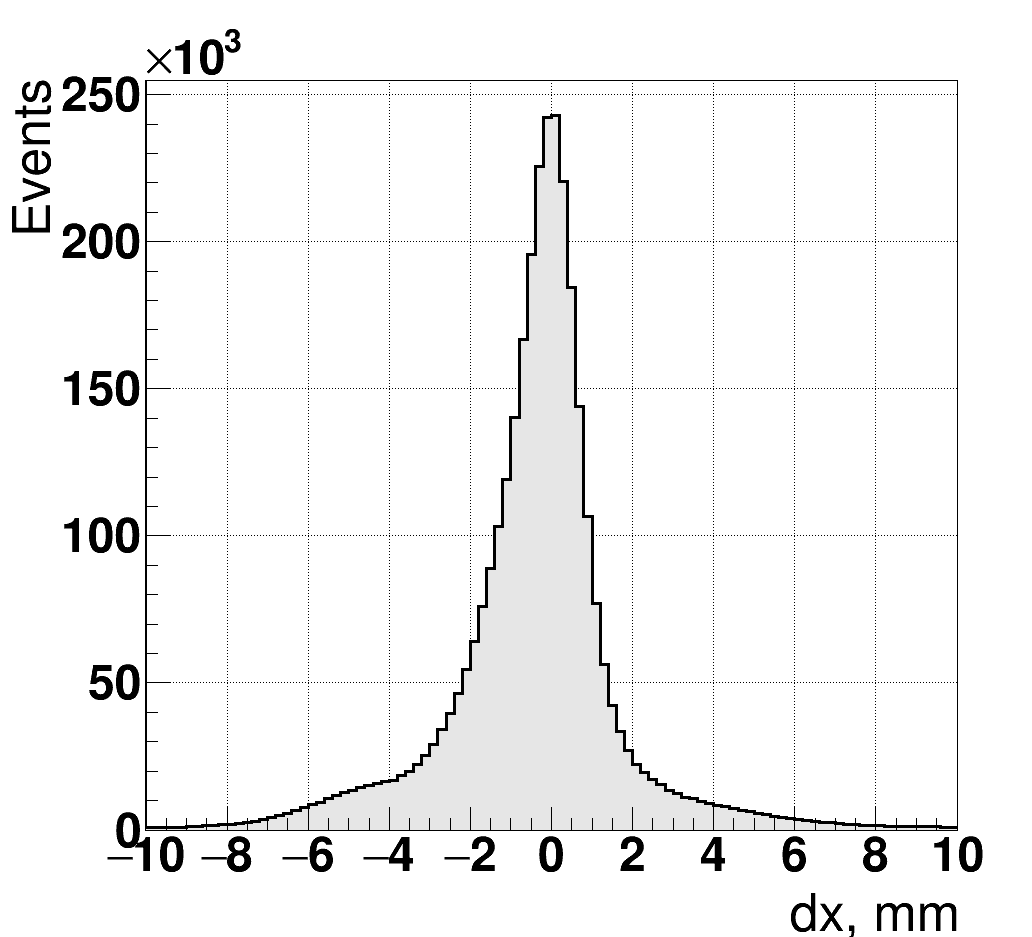}
\end{subfigure}
\hfill
\begin{subfigure}{0.45\textwidth}
    \includegraphics[width=\textwidth]{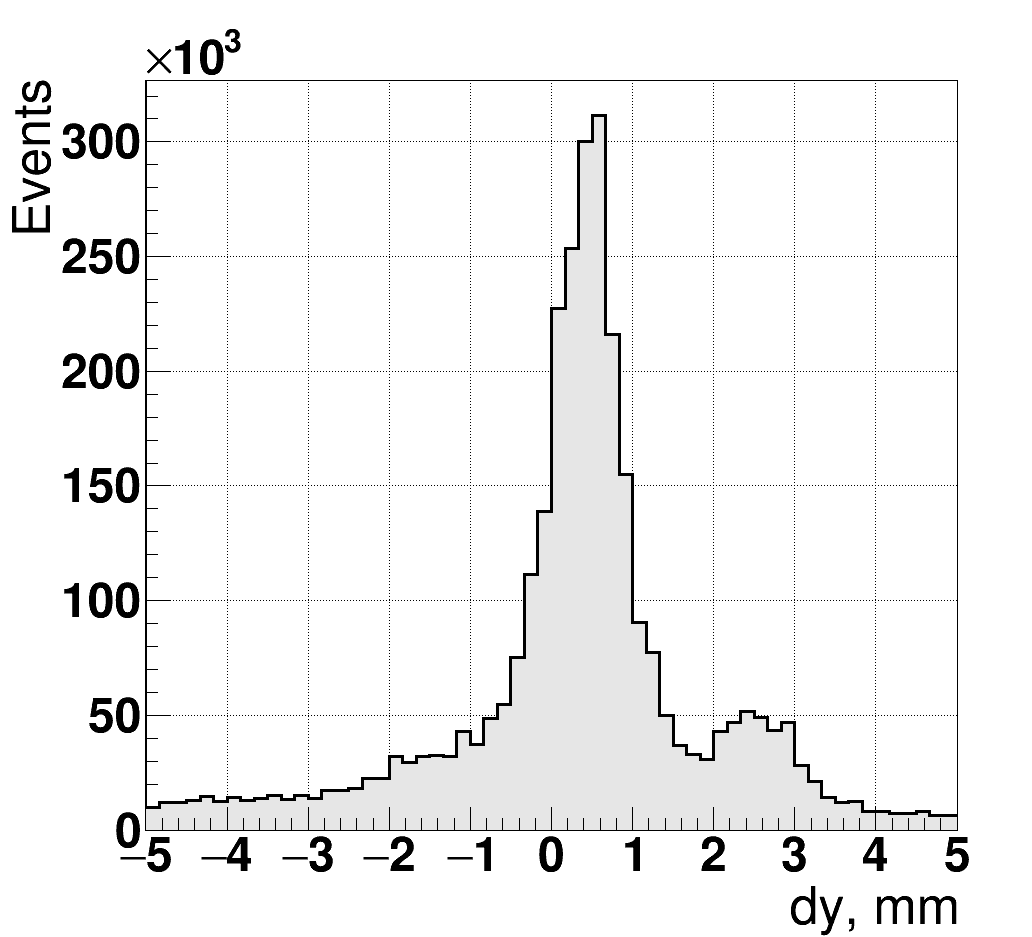}
\end{subfigure}
\caption{Histograms of the difference between the reconstructed position of the event and the position of the laser along the $x$-axis (\textbf{left}) and  $y$-axis (\textbf{right}) with the resolutions of 1.9 mm and 1.0 mm (FWHM), respectively.}
\label{fig:spatial_resolution}
\end{figure}

The photoelectron position is calculated for the $x$ and $y$-axes. For the coordinate perpendicular to the TLs orientation, corresponding to the $y$-axis, we compute the weighted average of the line $i$ that has the highest pulse with its two closest neighbors, as:
\begin{equation}
    Y_R=\frac{\sum _{k=i-1}^{i+1}y_k C_k}{\sum _{k=i-1}^{i+1} C_k}
\end{equation}
where $y_k$ is the $y$-coordinate of the central TL and $C_k$ the charge of the $k$-th TL. For the reconstruction on the $x$-axis, we computed the difference between the pulse time at both ends of the line multiplied by the signal propagation speed $s=(t_r-t_L)s/2$, where $t_R$ and $t_L$ are the times of the pulses measured at the right and left ends of the $i$-th line \cite{Follin_2022,Sung_2023}. The histograms presented in Fig. \ref{fig:spatial_resolution} show the spatial resolution along both axes over the detector's active surface. The computed values for the resolutions on the $x$ and $y$-axis are 1.9 mm and 1.0 mm (FWHM), respectively.
%%%%%%
%%%%%%%
 %%
\begin{figure}[ht!]
\centering
\includegraphics[width=0.5\textwidth]{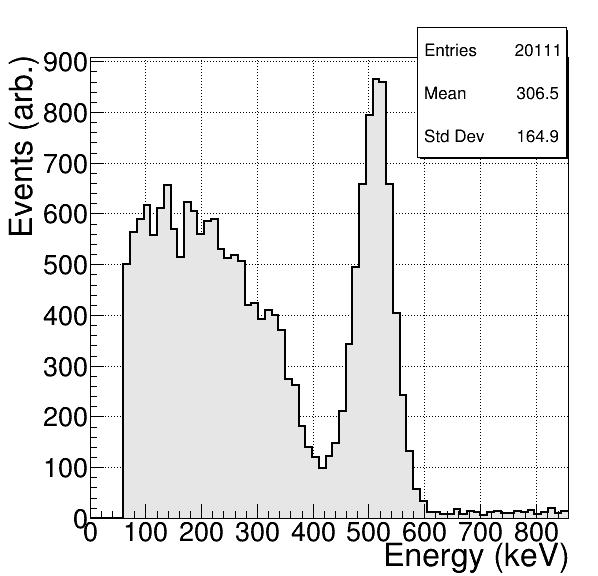}
\caption{The energy spectrum of the $^{22}$Na acquired with the reference LYSO/SiPM spectrometer.}
\label{fig:spectrum_SiPM}
\end{figure}
%%%

\section{CMP performances under gamma-ray radiation}
\label{sec:gamma_chapter}
\subsection{LYSO/SiPM reference spectrometer.}
\label{sec:spectrometer}
The spectrometer utilized for this test contains a 3 $\times$ 3 $\times$ 3 mm$^{3}$ LYSO:Ca,Ce co-doped crystal from Saint-Gobain (France) \cite{saint_gobain}, which features improved rise time, excellent decay time and light yield, and weaker afterglow as compared with LYSO:Ce \cite{Blahuta_2013}. The crystal was optically coupled with a 3 $\times$ 3 mm$^2$ Broadcom SiPM using histomount glue, diluted with 2/3 of Xylène, and wrapped in Teflon tape. 

A passive high-pass filter with a time constant of 2 ns was implemented at the SiPM output to remove the pulse slow time component. The time resolution measured with SAMPIC electronics is 103.5 $\pm$ 2 ps (FWHM). The signal was then amplified using a ZKL-1R5+ amplifier \cite{mini_circuits}, with a 40 dB gain, and bandwidth of 10 to 1500 MHz. After the amplification, the signal was attenuated with a passive attenuator of 3 dB model R411803124~\cite{radiall}, in order to match the dynamic range of the acquisition system.

We recorded spectrometer data for a $^{22}$Na source placed 10 cm away.  The SiPM was set at 18 \% overvoltage. Fig. \ref{fig:spectrum_SiPM} shows the measured energy spectrum, where we observe a well-defined photopeak at 511 keV preceded by the Compton edge.
The energy resolution measured on the photopeak is 13 \% (FWHM). The signal conditioning electronics are optimized for time performance rather than amplitude and the time window for the acquisition of the pulse does not cover the entire spectrometer pulse. The energy resolution computed in this context is excellent for the purpose of our study. The conditions for the acquisition are described in more detail in the next Section. 

\subsection{Experimental setup}
\label{sec:gamma_setup}
\begin{figure}[ht!]
\centering
\includegraphics[width=0.8\textwidth]{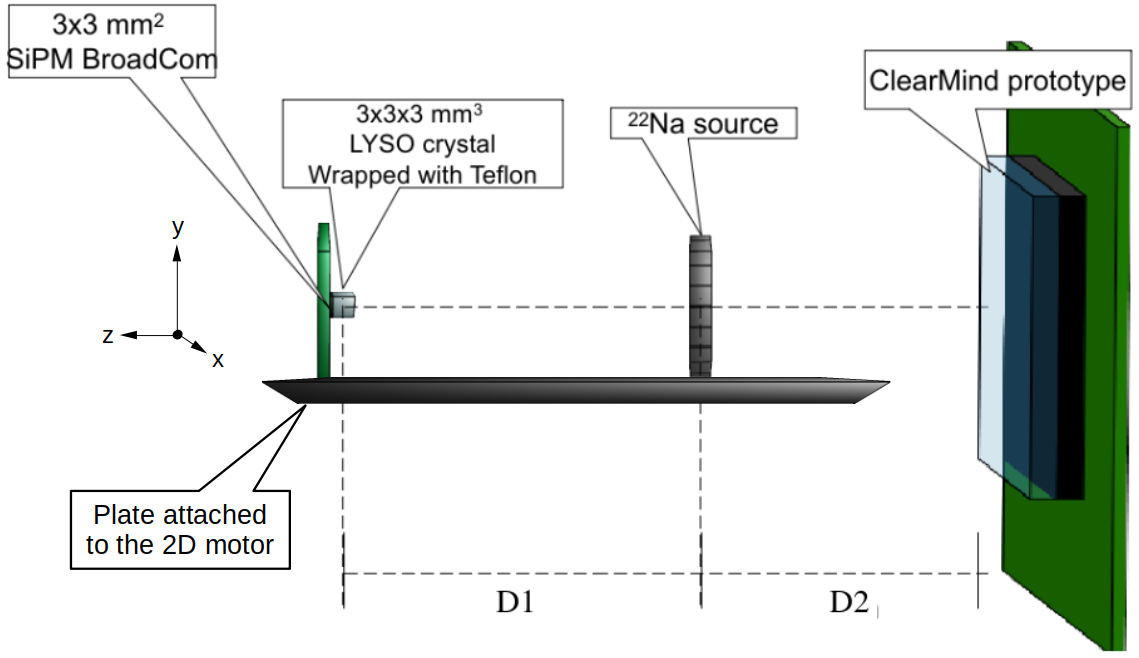}
\caption{\label{fig:gammas_setup} Scheme of the setup used for gamma measurements.}
\end{figure}

We took advantage of the test bench used for the photodetector calibration. A $^{22}$Na source was placed at a distance $D1$ from the CMP. The gamma spectrometer described in Section \ref{sec:spectrometer} was placed at a distance $D2$ from the source. Both, the spectrometer and the source, were mounted on a plate attached to the 2D motor station described in Section \ref{sec:calibration_setup} and aligned to the CMP. The LYSO/LYSO spectrometer and the gamma-ray source were moved together across the prototype's sensitive surface, allowing us to perform a 2D scan, as shown in Fig. \ref{fig:gammas_setup}. The boards' stack, which includes the transmission line board and the amplification boards, is the same as described in Section \ref{sec:calibration_setup}. Signals coming from the CMP and the spectrometer are recorded by a SAMPIC crate at 6.4 GS/s and stored on a PC.
%%%
%%%
Energy depositions in the reference spectrometer within the range 511 keV$\pm \Delta$, where $2\Delta_Energy =65.4$ keV is the FWHM of the photo-ionization peak, were considered for the analysis. The reconstruction and acquisition conditions will be described in the following Sections.

\subsection{Monte Carlo simulation}

   In order to understand the experimental results, we developed a detailed Monte Carlo modeling of the ClearMind detector \cite{Sung_2023}. 
    This simulation is based on the  
    Geant4 v7.0 \cite{agostinelli2003, Allison2006Feb, Allison2016Nov}. Furthermore, we developed dedicated software to model the photodetector, as well as analog and digital electronic components. The necessary parameters have been determined experimentally \cite{Follin_2021, Follin_2022}.
    The simulation of the detector response includes the following building blocks.
    \begin{enumerate}
        \item The gamma interaction in the crystal accounts for three processes: photoelectric conversion, Compton scattering, and Rayleigh diffusion. The first two processes produce relativistic electrons that emit optical
             photons from the photoelectric conversion of 511 keV gamma-rays through two mechanisms: Cherenkov radiation ($\sim$20 photons) and scintillation emission 
              ($\sim$165 photons). 

        \item Every optical photon is propagated individually by the simulation. During the propagation, all the main physical effects are taken into account: 
            photon absorption inside the PWO crystal,
              reflection or absorption on the crystal borders for the different types of the crystal surface
              (polished, absorbing), escape of photons from the crystal into the air. 

        \item Photocathode simulation includes modeling of the crystal/photocathode transmittance \cite{2018_Chen, Laurie2023}, absorption of photons by the photocathode, and extraction of generated photoelectrons as a function of the photon wavelength. 
        As a result, we produce 12 photoelectrons on average for a 511 keV $\gamma$-ray photoelectric conversion in the crystal with photocathode efficiency adjusted to the values measured in Section~\ref{sec:efficiency}.

        \item We then simulate the propagation and the multiplication of individual photoelectrons generated by the photocathode in the MCP-PMT and parametrize the main MCP-PMT response features: time response, MCP-PMT gain, gain fluctuation, and signal sharing between different output anodes.

        \item Finally, we simulate the signal readout through the transmission lines with realistic signal shapes, taking into account the possible overlay of several photoelectrons, electronics noise, and digitization sampling. 
              
    \end{enumerate}

Most of the simulation parameters (single-photoelectron transition time spread, MCP-PMT gain, gain fluctuation, signal sharing between lines, etc) are adjusted to the experimental results obtained from the characterization of the CMP using a pulsed laser in the single photoelectron regime. More details about the simulation can be found elsewhere \cite{Sung2022,Sung_2023,Cappellugola2021Oct}.

\subsection{CMP efficiency}
A high-efficiency detector is always advantageous, in particular for PET detectors \cite{Canot_2019}. For measuring the CMP's efficiency, we set $D1$ = 22 cm and $D2$ = 11 cm, according to the description of Fig. \ref{fig:gammas_setup}. The source and the spectrometer remained in a single position, close to the center of the detector surface. 
The incoming gamma-rays could interact within the crystal either by photoelectric effect, Compton scattering, or Rayleigh diffusion. According to previous simulations, when a 511 keV gamma-ray is converted via photoelectric effect, it produces mainly a 423 keV electron and X-rays that generates $\sim$ 187 optical photons in total \cite{Sung_2023}. By considering that the typical MCP-PMT photoelectron collection efficiency is 90 \% for these ALD-coated MCP-PMTs \cite{Lehmann2022}, we expect the CMP to be fully efficient at detecting photoelectron conversion. Interactions through the Compton effect generate electrons with energies below 340 keV. We have additional sensitivity for these events, particularly for the ones involving electrons with higher energies.

To estimate the detection efficiency, we used the \textit{Tag and Probe} method. The $^{22}$Na radioactive source emits pairs of antiparallel    511 keV annihilation photons and most of the time a 1.27 MeV gamma-ray. The SiPM-LYSO reference spectrometer carries out the \textit{tag} detection, triggers the acquisition of the SiPM spectrometers, and opens a 60 ns coincidence window allowing the acquisition of the pulses from the \textit{probe} CMP. The SiPM-LYSO reference spectrometer is fast enough to enable detector time resolution measurement below 200 ps (FWHM). The efficiency $\varepsilon$ on the probe detector is obtained as:
%%%%%%
\begin{equation}
\varepsilon = \frac{N_{probe}}{N_{tag}}
\end{equation}
%%%%
where $N_{tag}$ is the number of events recorded by the reference spectrometer, and $N_{probe}$ is the number of events recorded by the CMP.

We measured a detection efficiency \textbf{$\varepsilon$ = 28 $\pm$ 3 \%} (only statistical errors are reported). 

\subsection{CMP Spatial resolution}
\label{sec:spatial_response}
For this measurement, the spectrometer and the radioactive source were placed at $D1$ = 10 cm and $D2$ = 10 cm, according to Fig. \ref{fig:gammas_setup}. The acquisition was triggered by a coincidence between the CMP and the LYSO spectrometer within a time window of 20 ns. 
%%%%%%%%
\begin{figure}[h!]
\centering
\begin{subfigure}{0.48\textwidth}
    \includegraphics[width=\textwidth]{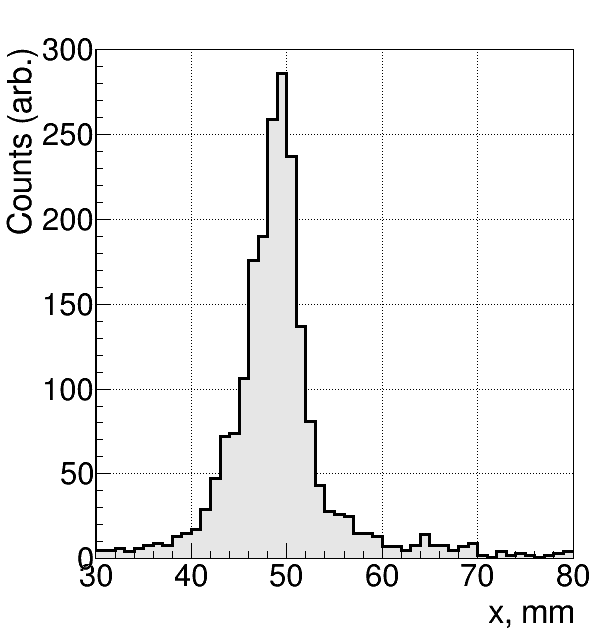}
\end{subfigure}
\hfill
\begin{subfigure}{0.48\textwidth}
    \includegraphics[width=\textwidth]{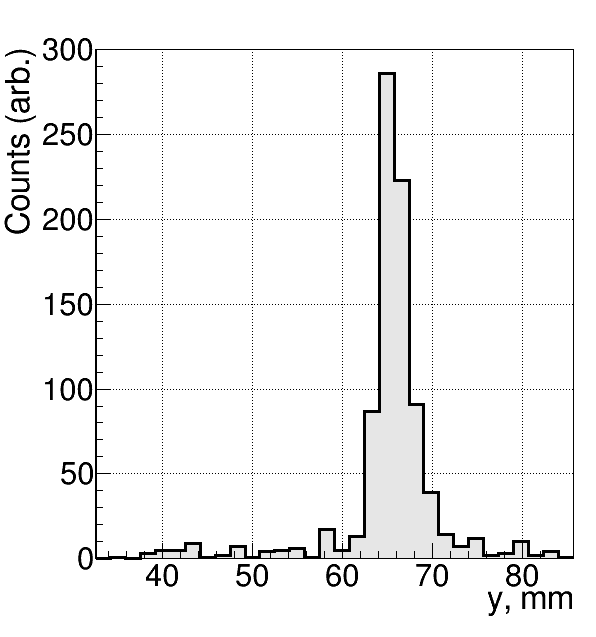}
\end{subfigure}
\caption{Reconstructed position of the gamma-ray source on the CMP: \textbf{Left :} 
 along the TLs, FWHM is 5.4 mm, and \textbf{Right :} across the TLs, FWHM is  3.5 mm.}
\label{fig:spatial_response_CTR}
\end{figure}
%%%%%%%%%%
%%%%%%%%
\begin{figure}[h!]
\centering
\begin{subfigure}{0.48\textwidth}
    \includegraphics[width=\textwidth]{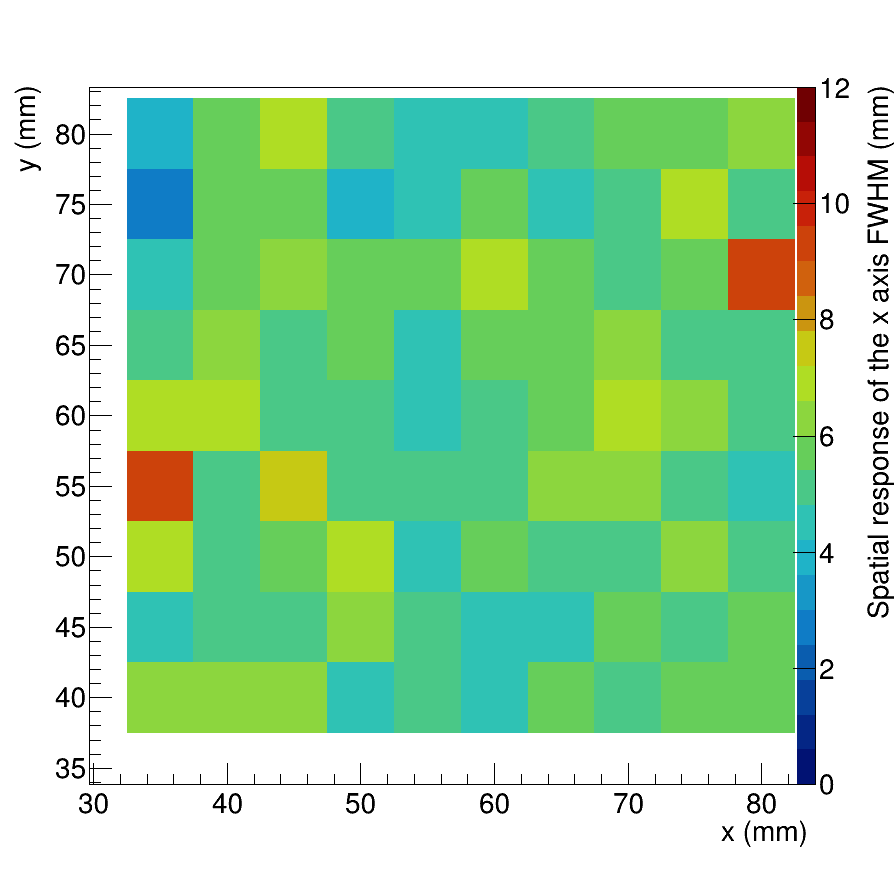}
\end{subfigure}
\hfill
\begin{subfigure}{0.48\textwidth}
    \includegraphics[width=\textwidth]{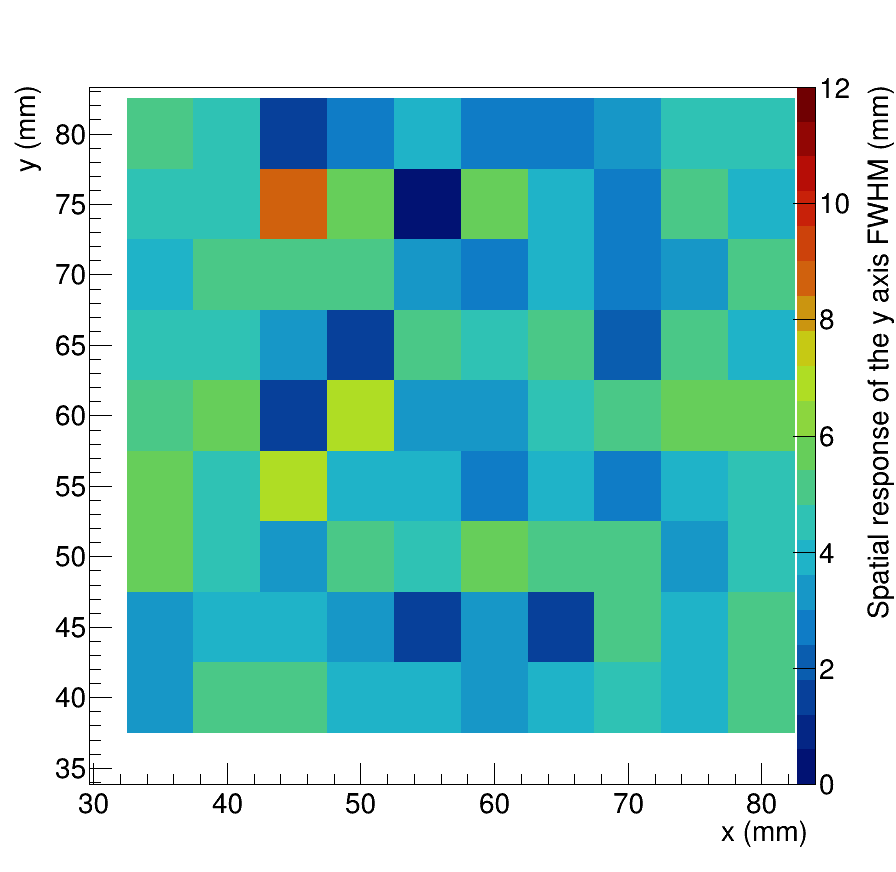}
\end{subfigure}1       
\caption{Spatial resolution (FWHM) of beam position reconstruction as a function of the (x, y) coordinates of the beam. 
\textbf{Left :} $x$-axis, along the transmission lines, and \textbf{Right :} $y$-axis, perpendicular to the transmission lines.}
\label{fig:spatial_response_scan}
\end{figure}
%%%%%%%%%%
The motor stage scanned the surface of the CMP by steps of 5 cm along both axes. We acquired a 60-minute run per data point. %The data acquisition rate was about 100 coincidences/s before events selection. 
The small solid angle of the reference spectrometer induces a trigger rate of $\sim$18 events/s. The coincidence condition with the SiPM spectrometer allows us to select 511 keV gamma-rays impinging in spot sizes of $\sim$3 mm diameter on the surface of the CMP. For the reconstruction of the interaction position, we utilized the algorithms described in Section \ref{sec:positionReconstruction}. The typical $x$ and $y$ distributions reconstructed from a fixed position ($x$ = 50 mm, $y$ = 65 mm), are shown in Fig. \ref{fig:spatial_response_CTR}. The spatial resolution measured for the $x$-axis parallel to the TLs is 5.4 mm (FWHM) and the spatial resolution measured for the $y$-axis perpendicular to the TLs is 3.5 mm (FWHM). Fig.~\ref{fig:spatial_response_scan} shows the spatial resolution as a function of the source position. We observe a uniform distribution of the spatial resolution along the $x$-axis, with a computed mean of 5.2 mm FWHM on the surface. 
%As the reconstruction depends on the time resolution along this axis, the spatial resolution can be improved if the time resolution improves.
We computed a mean of 4 mm FWHM for the spatial resolution for the $y$-axis, with a reasonable uniformity.

\subsection{CMP time response} 
\label{sec:time_response}
 To measure the detector time resolution, we used the same data collection from the setup described in Section \ref{sec:spatial_response}. %Whereas the event selection is within the 511 keV range at the spectrometer, we find events on the CMP with a variety in the number of channels above the threshold. 
%%%%%%%
\begin{figure}[h!]
\centering
\begin{subfigure}{0.52\textwidth}
    \includegraphics[width=\textwidth]{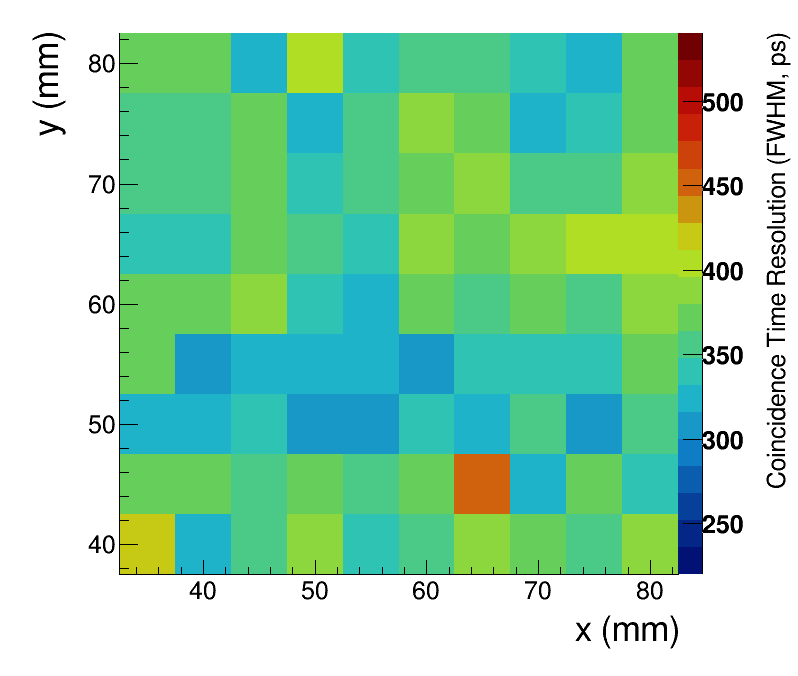}
\end{subfigure}
\hfill
\begin{subfigure}{0.45\textwidth}
    \includegraphics[width=\textwidth]{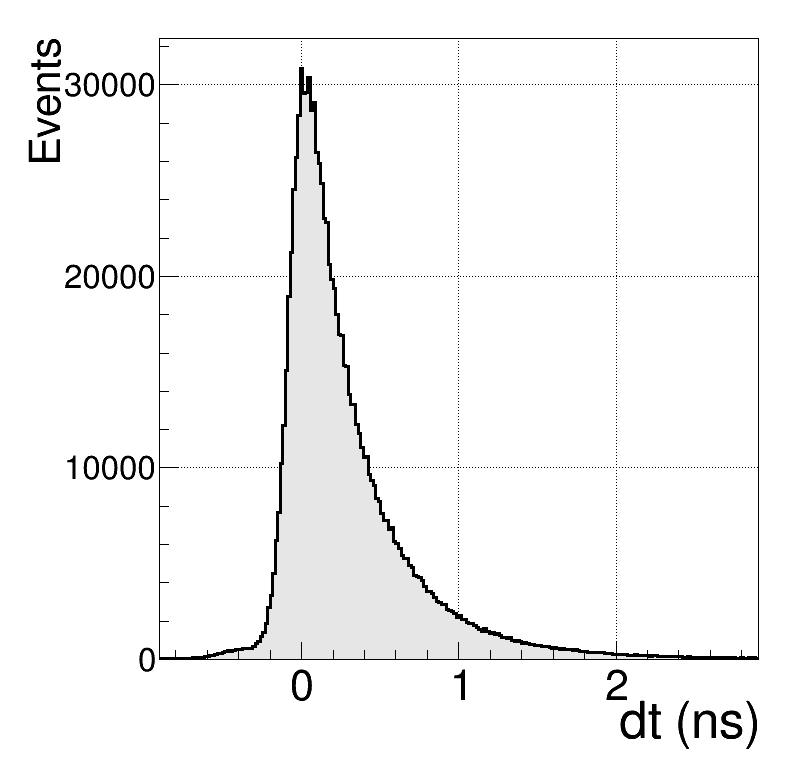}
\end{subfigure}
\caption{Measurements of the time distributions: \textbf{Left :} distribution of FWHM CTR as a function of the source position. \textbf{Right :} Time difference between the CMP and reference detector averaged over all the source positions. The FWHM is 370 ps.}
\label{fig:time_response_CTR}
\end{figure}
%%%%%%%
We selected events with 5 or more triggered lines, primarily arising from photoionization events.
The distribution of the time difference between the spectrometer signal and the fastest pulse recorded in coincidence for the CMP is shown in Fig. \ref{fig:time_response_CTR}. 
We observe a consistent CTR response for the majority of source positions, with typical median values of 350 ps, corresponding to a prototype resolution (computed after LYSO spectrometer contribution subtraction) of 330 ps. The time difference distribution averaged over all the source positions is shown in Fig. \ref{fig:time_response_CTR} on the right with a  CTR of 370 ps.

\section{Discussion}
\label{sec:discusion}
This study presents the first successful attempt to improve photon extraction probability from a PWO crystal through the direct deposition of a bialkali photocathode onto its surface. 
The feasibility of bialkali deposition on PWO proved to be challenging. After optimization of the passivation layer \cite{Yvon_2020,Follin_2021} between the crystal and the photocathode, a NPDE of 18 \% at 400 nm was demonstrated. The test cell presented in Section \ref{sec:testcell} was produced in 2021, and showed no significant efficiency variations over time. Moreover, this value is close to the value measured in the CMP and the value reported on \cite{Laurie2023}.

%%%
\begin{figure}[ht!]
\centering
\includegraphics[width=0.5\textwidth]{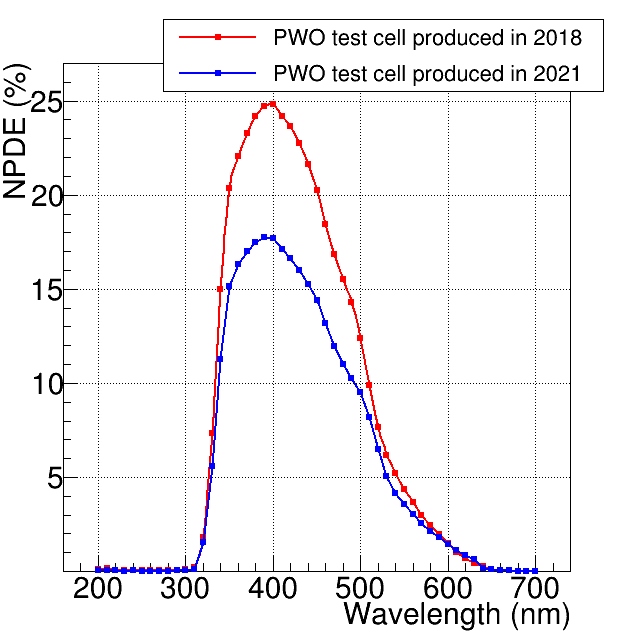}
\caption{NPDE comparison. The red curve corresponds to the NPDE measured on a PWO test cell produced in 2018, and the blue curve corresponds to the NPDE measured on a PWO test cell produced in 2021.}
\label{fig:QE_comparison}
\end{figure}
%%%

The CMP, which is described in Section~\ref{sec:detector}, serves as a compelling demonstration of the feasibility of this technique for constructing a scanner-sized detection module. Regrettably, this prototype exhibits limited optical photon detection efficiency, as presented in Section~\ref{sec:deteff}. As mentioned earlier, the test cell exhibits long-term photocathode stability. Moreover, a test cell produced in 2018 with a similar photocathode configuration, performs an NPDE of 25 \% at 400 nm, as shown in Fig. \ref{fig:QE_comparison}. This test cell presented stability over 4 years. Unfortunately, Photek Ltd. has not been able to reproduce a CMP with this NPDE until now. Nevertheless, future hardware efforts will focus on achieving 25 \% NPDE or more for future prototypes and improving the passivation layer, hence improving the optical light collection efficiency and the global detector performances.

%The gain value distribution across the CMP surface remains with the same pattern after multiple pressure adjustments between the anode pad and the transmission lines. Therefore, the pattern is not related directly to electrical couplings but is intrinsic to this prototype. This non-uniformity does not restrain a proper event reconstruction. Nevertheless, pressure adjustments on the Shin-Etsu rubber were found to impact the signal speed at the TLs, which need to be calibrated for event reconstruction. This motivates us to find an alternative electrical coupling method for mitigating systematics for future prototype versions.

We estimated the photoelectric interaction probability of a gamma-ray within the 5~mm PWO crystal of about 21.5~\%. Simulations showed that, from the expected 187 optical photons generated for a photoelectric interaction of a 511 keV gamma-ray, approximately 85 photons are absorbed by the photocathode and around 13 photons are converted into photoelectrons \cite{Sung_2023}. With this amount of photoelectrons and the high MCP-PMT collection efficiency, we are fully efficient for detecting gamma-rays interacting via the photoelectric effect.
The high Compton interaction probability of 25.2~\% for 511 keV gamma-rays passing through the 5~mm PWO crystal contributes to the total detection efficiency, which was measured to be $28 \pm 3$~\%. This estimation is consistent with the estimation from simulations, of 28.6~\%. By increasing the PWO crystal thickness to 10 mm, the photoelectric interaction probability would increase to 38.8~\%,  while the Compton interaction probability would rise to 44~\%, leading to an overall efficiency above 40~\%.

The reconstruction of the source position presented in this paper was computed using basic algorithms that demonstrate the possibility of reconstructing the source position with the proposed readout scheme. The collaboration is continuously working on optimizing the reconstruction of the interaction position including depth of interaction using neural networks. These methods demonstrated a great potential for improving the reconstruction of the interaction positions on Monte Carlo simulated events  \cite{Daniel_2023}. The improved version of the ClearMind detection module will incorporate an additional photosensor layer, such as a SiPM array, on the open front side of the crystal. This addition will enable us to enhance detection efficiency and, more importantly, to achieve high-precision reconstruction of the depth of interaction, as described in \cite{Yvon_2020}.

With the ClearMind detection paradigm, achieving precise timing is the ultimate goal, which could be reached through the detection of Cherenkov photons. Many factors contribute to the time resolution of our detector. The main one consists of :
 \begin{itemize}
    \item Time delay in optical photon production. Cherenkov photons are produced within a few ps time scale whereas scintillation photons are produced with time constants of 2 and 6 ns in PWO. Time information is therefore mainly provided by the detected Cherenkov photons which are rare.
    \item Random delay due to randomness in the optical photon collection path.
    \item Photo-detection processes in MCP-PMT, as documented in section \ref{sec:time_resolution}.
    \item Efficiency of the algorithm for time reconstruction from recorded data. Detected photon pulses pile up on readout lines, at random positions and random time delay. We use a modified Constant Fraction Discriminator algorithm in order to extract the time of the first optical photon detected in the event.
\end{itemize}
About the CMP, from the measured inputs used in a detailed Monte Carlo simulation described in ~\cite{Sung_2023}, the dominating contribution is the time delay in optical photon production. We demonstrated that 45~\% of the  Cherenkov photons generated from photoelectron conversion reach the photocathode. From the NPDE measured for the CMP, we computed that 0.7  Cherenkov photons are detected on average by applying a selection on the number of lines. This value corresponds to a probability of 48~\% for an event to contain at least one detected "best timing" Cherenkov photon, except in the case of a Compton interaction. This modest value limits the time resolution achieved with the CMP to 330~ps FWHM.

In refs. \cite{Kwon2021, Chen2024} excellent time resolutions (50 ps FWHM or better) were achieved in detecting 511-keV photons, using MCP-PMT pure Cherenkov radiator crystals. With pure Cherenkov radiator crystals, no "slow" time constant scintillation light is produced and only pure Cherenkov events enter the time distribution histogram. Currently, these excellent temporal performances have been achieved at the expense of significant efficiency loss, without provisions for measuring the deposited energy, and the spatial resolution of the detector has not been addressed in the publications.\newline
The CMP is designed to compromise between the specifications required for a TOF-PET detector module. Enhancing the Cherenkov light detection efficiency of CMP is mandatory to achieve a time resolution better than 100 ps. This is our first priority. We are actively working on improving the photocathode efficiency. Additionally, we plan to equip the front face of the crystal with a thin detection layer. Those upgrades will boost the number of photons detected and improve the overall performance of the ClearMind Detector design. Time reconstruction algorithms based on machine learning are an alternative we are investigating in detail.

\section{Conclusions}
In this study, we investigate the possibility of improving the timing performance of the PET detection module using Cherenkov photons while optimizing the optical interface by direct deposition of the photocathode on the PWO crystal.
 We proved the feasibility of direct photocathode deposition on a PWO crystal, with an efficiency of 18~\%, and demonstrated its stability over time.

We succeeded in creating a fully functional detection module. 
This prototype has an excellent single photoelectron time resolution of 70~ps FWHM all over its surface, spatial resolutions of 1.9 mm FWHM for the \textit{x}-axis parallel to the TLs, and 1.0 mm FWHM in the perpendicular direction. 
We tested the ClearMind prototype performance for detecting 511 keV gamma-rays and demonstrated the capability to reconstruct the source position, with a spatial resolution of 4 to 5 mm FWHM, using a simple reconstruction algorithm. This work is carried on by developing artificial intelligence reconstruction in order to achieve better resolutions, as well as for the estimation of the gamma-ray depth of interaction.
 CMP has non-optimal photocathode quantum efficiency, which limits the time resolution to 330~ps. The ClearMind technology will be upgraded in the future, as described in Section \ref{sec:discusion}, in order to improve the performance of the detection module.

%\appendix
%\section{Appendix}
%Please always give a title also for appendices.

\acknowledgments
The authors would like to thank Saint-Gobain Crystals for kindly providing the LYSO:Ce:Ca single crystals used to build reference spectrometers \cite{saint_gobain}.
We are grateful for the support and seed funding from the
CEA, Programme Exploratoire Bottom-Up, under grant No. 17P103-
CLEAR-MIND, and the French National Research Agency under the grant
No. ANR-19-CE19-0009-01.
Chi-Hsun Sung, Ph.D. is supported by the CEA NUMERICS program,
which has received funding from the European Union’s Horizon 2020
research and innovation program under the Marie Sklodowska-Curie
grant agreement No. 800945 - (NUMERICS - H2020-MSCA-COFUND-
2017).

%This is the most common positions for acknowledgments. A macro is
%available to maintain the same layout and spelling of the heading.

%\paragraph{Note added.} This is also a good position for notes added
%after the paper has been written.


%\begin{thebibliography}{99}
% Please avoid comments such as "For a review'', "For some examples",
% "and references therein" or move them in the text. In general,
% please leave only references in the bibliography and move all
% accessory text in footnotes.

% Also, please have only one work for each \bibitem.

%\end{thebibliography}
\bibliographystyle{unsrtnat}
\bibliography{biblio}{}

\begin{thebibliography}{44}
\providecommand{\natexlab}[1]{#1}
\providecommand{\url}[1]{\texttt{#1}}
\expandafter\ifx\csname urlstyle\endcsname\relax
  \providecommand{\doi}[1]{doi: #1}\else
  \providecommand{\doi}{doi: \begingroup \urlstyle{rm}\Url}\fi

\bibitem[Jiang et~al.(2019)Jiang, Chalich, and Deen]{Jiang2019SensorsFP}
Wei Jiang, Yamn Chalich, and M.~Jamal Deen.
\newblock Sensors for positron emission tomography applications.
\newblock \emph{Sensors (Basel, Switzerland)}, 19, November 2019.
\newblock URL \url{https://doi.org/10.3390%2Fs19225019}.

\bibitem[Ullah et~al.(2016)Ullah, Pratiwi, Cheon, Choi, and
  Yeom]{Ullah2016InstrumentationFT}
Muhammad~Nasir Ullah, Eva Pratiwi, Jimin Cheon, Hojong Choi, and J.~Y. Yeom.
\newblock Instrumentation for time-of-flight positron emission tomography.
\newblock \emph{Nuclear Medicine and Molecular Imaging}, 50:\penalty0 112--122,
  2016.
\newblock URL \url{https://doi.org/10.1007%2Fs13139-016-0401-5}.

\bibitem[Snyder et~al.(1981)Snyder, Thomas, and Ter-Pogossian]{Snyder1981}
Donald~L. Snyder, Lewis~J. Thomas, and Michel~M. Ter-Pogossian.
\newblock A mathematical model for positron-emission tomography systems having
  time-of-flight measurements.
\newblock \emph{IEEE Transactions on Nuclear Science}, 28\penalty0
  (3):\penalty0 3575--3583, 1981.
\newblock \doi{10.1109/TNS.1981.4332168}.
\newblock URL \url{https://epaper.kek.jp/p81/PDF/PAC1981_3575.PDF}.

\bibitem[Budinger(1983)]{Budinger1983}
Thomas Budinger.
\newblock Time-of-flight positron emission tomography: status relative to
  conventional pet.
\newblock \emph{Journal of nuclear medicine : official publication, Society of
  Nuclear Medicine}, 24 1:\penalty0 73--8, 1983.
\newblock URL \url{https://api.semanticscholar.org/CorpusID:671110}.

\bibitem[Lecoq(2017)]{8049484}
P.~Lecoq.
\newblock Pushing the limits in time-of-flight pet imaging.
\newblock \emph{IEEE Transactions on Radiation and Plasma Medical Sciences},
  1\penalty0 (6):\penalty0 473--485, 2017.
\newblock URL \url{https://doi.org/10.1109/TRPMS.2017.2756674}.

\bibitem[Lecoq et~al.(2020)Lecoq, Morel, Prior, Visvikis, Gundacker, Auffray,
  Kri{\v z}an, Turtos, Thers, Charbon, Varela, {De La Taille}, Rivetti, Breton,
  Pratte, Nuyts, Surti, Vandenberghe, Marsden, Parodi, Benlloch, and
  Benoit]{Lecoq2020}
Paul Lecoq, Christian Morel, {John O.} Prior, Dimitris Visvikis, Stefan
  Gundacker, Etiennette Auffray, Peter Kri{\v z}an, {Rosana Martinez} Turtos,
  Dominique Thers, Edoardo Charbon, Joao Varela, Christophe {De La Taille},
  Angelo Rivetti, Dominique Breton, {Jean Fran{\c c}ois} Pratte, Johan Nuyts,
  Suleman Surti, Stefaan Vandenberghe, Paul Marsden, Katia Parodi, {Jose Maria}
  Benlloch, and Mathieu Benoit.
\newblock Roadmap toward the 10 ps time-of-flight pet challenge.
\newblock \emph{Physics in Medicine and Biology}, 65\penalty0 (21), November
  2020.
\newblock \doi{10.1088/1361-6560/ab9500}.
\newblock URL
  \url{https://iopscience.iop.org/article/10.1088/1361-6560/ab9500}.

\bibitem[van Sluis et~al.(2019)van Sluis, de~Jong, Schaar, Noordzij, van Snick,
  Dierckx, Borra, Willemsen, and Boellaard]{vanSluis2019PerformanceCO}
Joyce van Sluis, J.~de~Jong, Jenny Schaar, Walter Noordzij, Paul~J.H. van
  Snick, Rudi~A. Dierckx, Ronald J.~H. Borra, Antoon T.~M. Willemsen, and
  Ronald Boellaard.
\newblock Performance characteristics of the digital biograph vision pet/ct
  system.
\newblock \emph{The Journal of Nuclear Medicine}, 60:\penalty0 1031 -- 1036,
  2019.
\newblock URL \url{https://doi.org/10.2967/jnumed.118.215418}.

\bibitem[Yvon et~al.(2020)Yvon, Sharyy, Follin, Bard, Breton, Maalmi, Morel,
  and Delagnes]{Yvon_2020}
D.~Yvon, V.~Sharyy, M.~Follin, J.P. Bard, D.~Breton, J.~Maalmi, C.~Morel, and
  E.~Delagnes.
\newblock {Design study of a ``scintronic'' crystal targeting tens of
  picoseconds time resolution for gamma ray imaging: the ClearMind detector}.
\newblock \emph{{Journal of Instrumentation}}, 15\penalty0 (07):\penalty0
  P07029, 2020.
\newblock URL \url{https://hal.science/hal-02899246}.

\bibitem[Follin et~al.(2021)Follin, Sharyy, Bard, Korzhik, and
  Yvon]{Follin_2021}
M.~Follin, V.~Sharyy, J-P. Bard, M.~Korzhik, and D.~Yvon.
\newblock {Scintillating properties of today available lead tungstate
  crystals}.
\newblock \emph{{Journal of Instrumentation}}, 16\penalty0 (08):\penalty0
  P08040, 2021.
\newblock \doi{10.1088/1748-0221/16/08/P08040}.
\newblock URL \url{https://hal.science/hal-03197498}.

\bibitem[CRYTUR(2023)]{crytur}
CRYTUR.
\newblock spol. s r.o. {N}a {L}ukách 2283 51101 {T}urnov {C}zech {R}epublic,
  2023.
\newblock URL \url{https://www.crytur.com}.

\bibitem[Ltd.(2023)]{photek_MAPMT253}
Photek Ltd.
\newblock {MAPMT}-253 {M}ulti-{A}node {MCP-PMT} {D}atasheet, 2023.
\newblock URL
  \url{https://www.photek.com/pdf/datasheets/detectors/Auratek-Square-Detector-13SEP17.pdf}.

\bibitem[Berger et~al.(2010)Berger, Hubbell, Seltzer, Coursey, and
  Zucker]{XCOM}
Martin Berger, J~Hubbell, Stephen Seltzer, J~Coursey, and D~Zucker.
\newblock Xcom: Photon cross section database, 2010.
\newblock URL \url{http://physics.nist.gov/xcom}.

\bibitem[Follin et~al.(2022)Follin, Chyzh, Sung, Breton, Maalmi, Chaminade,
  Delagnes, Schäfers, Weinheimer, Yvon, and Sharyy]{Follin_2022}
M.~Follin, R.~Chyzh, C.-H. Sung, D.~Breton, J.~Maalmi, T.~Chaminade,
  E.~Delagnes, K.~Schäfers, C.~Weinheimer, D.~Yvon, and V.~Sharyy.
\newblock High resolution {MCP}-{PMT} readout using transmission lines.
\newblock \emph{Nuclear Instruments and Methods in Physics Research Section A:
  Accelerators, Spectrometers, Detectors and Associated Equipment},
  1027:\penalty0 166092, mar 2022.
\newblock URL \url{https://doi.org/10.1016/j.nima.2021.166092}.

\bibitem[SHIN-ETSU(2023)]{interconnector}
SHIN-ETSU.
\newblock {MT-T}ype of {I}nter-{C}onnector {D}atasheet, 2023.
\newblock URL
  \url{https://www.shinetsu.info/product/mt-type-of-inter-connector/?attachment_id=1687&download_file=d2bw67wo7s226}.

\bibitem[Breton(March 2014)]{BRETON2014}
Dominique Breton.
\newblock {Measuring time with a 5-ps precision at the systel level with the
  WaveCatcher family of SCA-based fast digitizers}.
\newblock In \emph{{Workshop in picosecond photon sensors for physics and
  medical application}}, March 2014.
\newblock URL \url{https://indico.cern.ch/event/306859/session/3/contribution/
  9/material/slides/1.pdf}.

\bibitem[Delagnes et~al.(March 2014)Delagnes, Grabas, Breton, and
  Maalmi]{Delagnes2014}
E.~Delagnes, H.~Grabas, D.~Breton, and J.~Maalmi.
\newblock {The sampic WTDC chip}.
\newblock \emph{Workshop on Picosecond Photon Sensors for physics and medical
  application}, March 2014.
\newblock URL
  \url{https://indico.cern.ch/event/306859/contributions/705887/attachments/584423/804477/ED_SAMPIC_Clermont_mars2014.pdf}.

\bibitem[Delagnes et~al.(2015)Delagnes, Breton, Grabas, Maalmi, and
  Rusquart]{Delagnes:2015oda}
E.~Delagnes, D.~Breton, H.~Grabas, J.~Maalmi, and P.~Rusquart.
\newblock {Reaching a few picosecond timing precision with the 16-channel
  digitizer and timestamper SAMPIC ASIC}.
\newblock \emph{Nucl. Instrum. Meth. A}, A787:\penalty0 245--249, 2015.
\newblock \doi{10.1016/j.nima.2014.12.042}.
\newblock URL
  \url{https://www.sciencedirect.com/science/article/abs/pii/S0168900214014843}.

\bibitem[Breton et~al.(2016)Breton, De~Cacqueray, Delagnes, Grabas, Maalmi,
  Minafra, Royon, and Saimpert]{Breton2016}
D.~Breton, V.~De~Cacqueray, E.~Delagnes, H.~Grabas, J.~Maalmi, N.~Minafra,
  C.~Royon, and M.~Saimpert.
\newblock {Measurements of timing resolution of ultra-fast silicon detectors
  with the {SAMPIC} waveform digitizer}.
\newblock \emph{Nucl. Instrum. Meth. A}, 835\penalty0 (Supplement C):\penalty0
  51--60, 2016.
\newblock \doi{10.1016/j.nima.2016.08.019}.
\newblock URL
  \url{https://www.sciencedirect.com/science/article/pii/S0168900216308373}.

\bibitem[Breton et~al.(2020)Breton, Cheikali, Delagnes, Maalmi, Rusquart, and
  Vallerand]{Breton2020}
D.~Breton, C.~Cheikali, E.~Delagnes, J.~Maalmi, P.~Rusquart, and P.~Vallerand.
\newblock {Fast electronics for particle Time-Of-Flight measurement, with focus
  on the SAMPIC ASIC}.
\newblock \emph{Nuovo Cimento C}, 43\penalty0 (1):\penalty0 7, 2020.
\newblock \doi{10.1393/ncc/i2020-20007-6}.
\newblock URL \url{https://www.sif.it/papers/?pid=ncc12001}.

\bibitem[Motta and Schonert(2005)]{Motta2005a}
Dario Motta and Stefan Schonert.
\newblock {Optical properties of Bialkali photocathodes}.
\newblock \emph{Nucl. Instrum. Meth. A}, A539:\penalty0 217--235, 2005.
\newblock \doi{10.1016/j.nima.2004.10.009}.
\newblock URL
  \url{https://www.sciencedirect.com/science/article/abs/pii/S0168900204022132}.

\bibitem[Sung et~al.(2023)Sung, Cappellugola, Follin, Curtoni, Dupont, Morel,
  Galindo-Tellez, Chyzh, Breton, Maalmi, Yvon, and Sharyy]{Sung_2023}
C.-H. Sung, L.~Cappellugola, M.~Follin, S.~Curtoni, M.~Dupont, C.~Morel,
  A.~Galindo-Tellez, R.~Chyzh, D.~Breton, J.~Maalmi, D.~Yvon, and V.~Sharyy.
\newblock Detailed simulation for the clearmind prototype detection module and
  event reconstruction using artificial intelligence.
\newblock \emph{Nuclear Instruments and Methods in Physics Research Section A:
  Accelerators, Spectrometers, Detectors and Associated Equipment},
  1053:\penalty0 168357, August 2023.
\newblock \doi{https://doi.org/10.1016/j.nima.2023.168357}.
\newblock URL
  \url{https://www.sciencedirect.com/science/article/pii/S0168900223003479}.

\bibitem[Ocean~Optics(2023)]{DH_2000}
Inc. Ocean~Optics.
\newblock {DH-2000-BAL} {L}ight {S}ource, 2023.
\newblock URL
  \url{https://www.oceaninsight.com/products/light-sources/uv-vis-nir-light-sources/dh-2000-bal/?qty=1}.

\bibitem[Zolix Instruments~Co.(2023)]{monochromator}
Ltd. Zolix Instruments~Co.
\newblock Omni-$\lambda$200i, 2023.
\newblock URL \url{https://www.zolix.com.cn/en/prodcon_370_376_741.html}.

\bibitem[Instruments(2023)]{picoAmp}
Keithley Instruments.
\newblock Electrometer {M}odel 6517{B}, 2023.
\newblock URL
  \url{https://www.tek.com/en/datasheet/6517b-electrometer-high-resistance-meter}.

\bibitem[Thorlabs(2023)]{ref_thorlabs}
Inc. Thorlabs.
\newblock {FDS1010} {S}i detector {D}atasheet, 2023.
\newblock URL
  \url{https://www.thorlabs.com/drawings/51e19728c72ae1ee-36153490-E029-7FE0-DB9940D12A2CABE3/FDS1010-CAL-SpecSheet.pdf}.

\bibitem[Corporation.(2022)]{ref_newport}
Newport Corporation.
\newblock 818-{UV/DB} detector {D}atasheet, 2022.
\newblock URL
  \url{https://www.newport.com/medias/sys_master/images/images/hbd/h46/9599504875550/DS-051502-818-Series-Photodiode-Datasheet.pdf}.

\bibitem[K.K.(2017)]{hamamatsu}
Hamamatsu~Photonics K.K.
\newblock \emph{{PHOTOMULTIPLIER TUBES}, {B}asics and {A}pplications}.
\newblock Hamamatsu Photonics K.K. Electron Tube Division, 2017.
\newblock URL
  \url{https://www.hamamatsu.com/content/dam/hamamatsu-photonics/sites/documents/99_SALES_LIBRARY/etd/PMT_handbook_v4E.pdf}.

\bibitem[GmbH(2017)]{PiLas}
Advansed Laser Diode System~A.L.S. GmbH.
\newblock Picosecond diode laser - pilas, manual and test report cea
  {P}i{L040XSM}-1\_825, 2017.
\newblock URL
  \url{https://www.nktphotonics.com/products/pulsed-diode-lasers/pilas/}.

\bibitem[{Chen} et~al.(2018){Chen}, {Tian}, {Guo}, {Wei}, {Liu}, {Sai}, {Wang},
  {Lu}, {Wang}, {Wang}, {He}, {Tian}, {Xin}, and {Guo}]{2018_Chen}
Ping {Chen}, Jinshou {Tian}, Lehui {Guo}, Yonglin {Wei}, Hulin {Liu}, Xiaofeng
  {Sai}, Xing {Wang}, Yu~{Lu}, Chao {Wang}, Junfeng {Wang}, Kai {He}, Liping
  {Tian}, Liwei {Xin}, and Haitao {Guo}.
\newblock {Photoelectron backscattering in the microchannel plate
  photomultiplier tube}.
\newblock \emph{Nuclear Instruments and Methods in Physics Research Section A:
  Accelerators, Spectrometers, Detectors and Associated Equipment},
  912:\penalty0 112--114, December 2018.
\newblock \doi{10.1016/j.nima.2017.10.081}.
\newblock URL
  \url{https://www.sciencedirect.com/science/article/pii/S0168900217311701}.

\bibitem[sai(2023)]{saint_gobain}
{S}aint-{G}obain, {F}rance, 2023.
\newblock URL \url{https://www.crystals.saint-gobain.com}.

\bibitem[Blahuta et~al.(2013)Blahuta, Bessière, Viana, Dorenbos, and
  Ouspenski]{Blahuta_2013}
S.~Blahuta, A.~Bessière, B.~Viana, P.~Dorenbos, and V.~Ouspenski.
\newblock Evidence and consequences of {C}e $^{4+}$ in {LYSO:Ce,Ca} and
  {LYSO:Ce,Mg} single crystals for medical imaging applications.
\newblock \emph{IEEE Transactions on Nuclear Science}, 60\penalty0
  (4):\penalty0 3134--3141, 2013.
\newblock \doi{10.1109/TNS.2013.2269700}.
\newblock URL \url{https://doi.org/10.1109/TNS.2013.2269700}.

\bibitem[Mini-Circuits.(2023)]{mini_circuits}
Mini-Circuits.
\newblock Coaxial amplifier zkl-1r5+, datasheet, 2023.
\newblock URL \url{https://www.minicircuits.com/pdfs/ZKL-1R5.pdf}.

\bibitem[Radiall.(2023)]{radiall}
Radiall.
\newblock 3db attenuator, datasheet, 2023.
\newblock URL \url{https://www.farnell.com/datasheets/2092151.pdf}.

\bibitem[Agostinelli et~al.(2003)]{agostinelli2003}
S.~Agostinelli et~al.
\newblock {GEANT4: A Simulation toolkit}.
\newblock \emph{Nucl. Instrum. Meth. A}, 506:\penalty0 250--303, 2003.
\newblock \doi{10.1016/S0168-9002(03)01368-8}.
\newblock URL
  \url{https://www.sciencedirect.com/science/article/abs/pii/S0168900203013688?via%3Dihub}.

\bibitem[Allison et~al.(2006)Allison, Amako, Apostolakis, Araujo, Dubois, Asai,
  Barrand, Capra, Chauvie, Chytracek, Cirrone, Cooperman, Cosmo, Cuttone,
  Daquino, Donszelmann, Dressel, Folger, Foppiano, Generowicz, Grichine,
  Guatelli, Gumplinger, Heikkinen, Hrivnacova, Howard, Incerti, Ivanchenko,
  Johnson, Jones, Koi, Kokoulin, Kossov, Kurashige, Lara, Larsson, Lei, Link,
  Longo, Maire, Mantero, Mascialino, McLaren, Lorenzo, Minamimoto, Murakami,
  Nieminen, Pandola, Parlati, Peralta, Perl, Pfeiffer, Pia, Ribon, Rodrigues,
  Russo, Sadilov, Santin, Sasaki, Smith, Starkov, Tanaka, Tcherniaev, Tome,
  Trindade, Truscott, Urban, Verderi, Walkden, Wellisch, Williams, Wright, and
  Yoshida]{Allison2006Feb}
J.~Allison, K.~Amako, J.~Apostolakis, H.~Araujo, P.~Arce Dubois, M.~Asai,
  G.~Barrand, R.~Capra, S.~Chauvie, R.~Chytracek, G.~A.~P. Cirrone,
  G.~Cooperman, G.~Cosmo, G.~Cuttone, G.~G. Daquino, M.~Donszelmann,
  M.~Dressel, G.~Folger, F.~Foppiano, J.~Generowicz, V.~Grichine, S.~Guatelli,
  P.~Gumplinger, A.~Heikkinen, I.~Hrivnacova, A.~Howard, S.~Incerti,
  V.~Ivanchenko, T.~Johnson, F.~Jones, T.~Koi, R.~Kokoulin, M.~Kossov,
  H.~Kurashige, V.~Lara, S.~Larsson, F.~Lei, O.~Link, F.~Longo, M.~Maire,
  A.~Mantero, B.~Mascialino, I.~McLaren, P.~Mendez Lorenzo, K.~Minamimoto,
  K.~Murakami, P.~Nieminen, L.~Pandola, S.~Parlati, L.~Peralta, J.~Perl,
  A.~Pfeiffer, M.~G. Pia, A.~Ribon, P.~Rodrigues, G.~Russo, S.~Sadilov,
  G.~Santin, T.~Sasaki, D.~Smith, N.~Starkov, S.~Tanaka, E.~Tcherniaev,
  B.~Tome, A.~Trindade, P.~Truscott, L.~Urban, M.~Verderi, A.~Walkden, J.~P.
  Wellisch, D.~C. Williams, D.~Wright, and H.~Yoshida.
\newblock {Geant4 developments and applications}.
\newblock \emph{IEEE Trans. Nucl. Sci.}, 53\penalty0 (1):\penalty0 270--278,
  February 2006.
\newblock \doi{10.1109/TNS.2006.869826}.
\newblock URL \url{https://ieeexplore.ieee.org/document/1610988}.

\bibitem[Allison et~al.(2016)Allison, Amako, Apostolakis, Arce, Asai, Aso,
  Bagli, Bagulya, Banerjee, Barrand, Beck, Bogdanov, Brandt, Brown, Burkhardt,
  Canal, Cano-Ott, Chauvie, Cho, Cirrone, Cooperman,
  Cort{\ifmmode\acute{e}\else\'{e}\fi}s-Giraldo, Cosmo, Cuttone, Depaola,
  Desorgher, Dong, Dotti, Elvira, Folger, Francis, Galoyan, Garnier, Gayer,
  Genser, Grichine, Guatelli, Gu{\ifmmode\grave{e}\else\`{e}\fi}ye, Gumplinger,
  Howard,
  H{\ifmmode\check{r}\else\v{r}\fi}ivn{\ifmmode\acute{a}\else\'{a}\fi}{\ifmmode\check{c}\else\v{c}\fi}ov{\ifmmode\acute{a}\else\'{a}\fi},
  Hwang, Incerti, Ivanchenko, Ivanchenko, Jones, Jun, Kaitaniemi, Karakatsanis,
  Karamitros, Kelsey, Kimura, Koi, Kurashige, Lechner, Lee, Longo, Maire,
  Mancusi, Mantero, Mendoza, Morgan, Murakami, Nikitina, Pandola, Paprocki,
  Perl, Petrovi{\ifmmode\acute{c}\else\'{c}\fi}, Pia, Pokorski, Quesada, Raine,
  Reis, Ribon, Risti{\ifmmode\acute{c}\else\'{c}\fi}~Fira, Romano, Russo,
  Santin, Sasaki, Sawkey, Shin, Strakovsky, Taborda, Tanaka,
  Tom{\ifmmode\acute{e}\else\'{e}\fi}, Toshito, Tran, Truscott, Urban,
  Uzhinsky, Verbeke, Verderi, Wendt, Wenzel, Wright, Wright, Yamashita, Yarba,
  and Yoshida]{Allison2016Nov}
J.~Allison, K.~Amako, J.~Apostolakis, P.~Arce, M.~Asai, T.~Aso, E.~Bagli,
  A.~Bagulya, S.~Banerjee, G.~Barrand, B.~R. Beck, A.~G. Bogdanov, D.~Brandt,
  J.~M.~C. Brown, H.~Burkhardt, {\relax Ph}.~Canal, D.~Cano-Ott, S.~Chauvie,
  K.~Cho, G.~A.~P. Cirrone, G.~Cooperman, M.~A.
  Cort{\ifmmode\acute{e}\else\'{e}\fi}s-Giraldo, G.~Cosmo, G.~Cuttone,
  G.~Depaola, L.~Desorgher, X.~Dong, A.~Dotti, V.~D. Elvira, G.~Folger,
  Z.~Francis, A.~Galoyan, L.~Garnier, M.~Gayer, K.~L. Genser, V.~M. Grichine,
  S.~Guatelli, P.~Gu{\ifmmode\grave{e}\else\`{e}\fi}ye, P.~Gumplinger, A.~S.
  Howard,
  I.~H{\ifmmode\check{r}\else\v{r}\fi}ivn{\ifmmode\acute{a}\else\'{a}\fi}{\ifmmode\check{c}\else\v{c}\fi}ov{\ifmmode\acute{a}\else\'{a}\fi},
  S.~Hwang, S.~Incerti, A.~Ivanchenko, V.~N. Ivanchenko, F.~W. Jones, S.~Y.
  Jun, P.~Kaitaniemi, N.~Karakatsanis, M.~Karamitros, M.~Kelsey, A.~Kimura,
  T.~Koi, H.~Kurashige, A.~Lechner, S.~B. Lee, F.~Longo, M.~Maire, D.~Mancusi,
  A.~Mantero, E.~Mendoza, B.~Morgan, K.~Murakami, T.~Nikitina, L.~Pandola,
  P.~Paprocki, J.~Perl, I.~Petrovi{\ifmmode\acute{c}\else\'{c}\fi}, M.~G. Pia,
  W.~Pokorski, J.~M. Quesada, M.~Raine, M.~A. Reis, A.~Ribon,
  A.~Risti{\ifmmode\acute{c}\else\'{c}\fi}~Fira, F.~Romano, G.~Russo,
  G.~Santin, T.~Sasaki, D.~Sawkey, J.~I. Shin, I.~I. Strakovsky, A.~Taborda,
  S.~Tanaka, B.~Tom{\ifmmode\acute{e}\else\'{e}\fi}, T.~Toshito, H.~N. Tran,
  P.~R. Truscott, L.~Urban, V.~Uzhinsky, J.~M. Verbeke, M.~Verderi, B.~L.
  Wendt, H.~Wenzel, D.~H. Wright, D.~M. Wright, T.~Yamashita, J.~Yarba, and
  H.~Yoshida.
\newblock {Recent developments in Geant4}.
\newblock \emph{Nucl. Instrum. Meth. A}, 835:\penalty0 186--225, 2016.
\newblock \doi{10.1016/j.nima.2016.06.125}.
\newblock URL
  \url{https://www.sciencedirect.com/science/article/pii/S0168900216306957}.

\bibitem[Cappellugola(2023)]{Laurie2023}
Laurie Cappellugola.
\newblock Modélisation monte carlo d'un détecteur scintronique à haute
  résolution spatio-temporelle couplé à un tube multiplicateur à galette de
  micro-canaux, phd thesis, 2023.
\newblock URL \url{https://theses.fr/s301206}.

\bibitem[Sung(2022)]{Sung2022}
Chi-Hsun Sung.
\newblock \emph{Simulation and artificial intelligence for a gamma detector for
  high resolution PET imaging}.
\newblock PhD thesis, Université Paris-Saclay, 2022.
\newblock URL \url{http://www.theses.fr/2022UPASP060}.

\bibitem[Cappellugola et~al.(2021)Cappellugola, Curtoni, Dupont, Sung, Sharyy,
  Thibault, Yvon, and Morel]{Cappellugola2021Oct}
L.~Cappellugola, S.~Curtoni, M.~Dupont, C.-H. Sung, V.~Sharyy, C.~Thibault,
  D.~Yvon, and C.~Morel.
\newblock {Modelisation of Light Transmission through Surfaces with Thin Film
  Optical Coating in Geant4}.
\newblock In \emph{{2021 IEEE Nuclear Science Symposium and Medical Imaging
  Conference (NSS/MIC)}}, pages 1--5. IEEE, 2021.
\newblock \doi{10.1109/NSS/MIC44867.2021.9875513}.
\newblock URL \url{https://ieeexplore.ieee.org/document/9875513}.

\bibitem[Canot et~al.(2019)Canot, Alokhina, Abbon, Bard, Breton, Delagnes,
  Maalmi, Tauzin, Yvon, and Sharyy]{Canot_2019}
C.~Canot, M.~Alokhina, P.~Abbon, J.P. Bard, D.~Breton, E.~Delagnes, J.~Maalmi,
  G.~Tauzin, D.~Yvon, and V.~Sharyy.
\newblock {Fast and efficient detection of 511 keV photons using Cherenkov
  light in PbF$_2$ crystal, coupled to a MCP-PMT and SAMPIC digitization
  module}.
\newblock \emph{Journal of Instrumentation}, 14\penalty0 (12):\penalty0 P12001,
  dec 2019.
\newblock \doi{10.1088/1748-0221/14/12/P12001}.
\newblock URL \url{https://dx.doi.org/10.1088/1748-0221/14/12/P12001}.

\bibitem[Lehmann et~al.(2022)Lehmann, Böhm, Miehling, Pfaffinger, Stelter,
  Uhlig, Ali, Belias, Dzhygadlo, Gerhardt, Krebs, Lehmann, Peters, Schepers,
  Schwarz, Schwiening, Traxler, L.~Schmitt~d, Etzelmüller, Föhl, Hayrapetyan,
  Kreutzfeld, Rieke, Schmidt, Wasem, and Sfienti]{Lehmann2022}
A.~Lehmann, M.~Böhm, D.~Miehling, M.~Pfaffinger, S.~Stelter, F.~Uhlig, A.~Ali,
  A.~Belias, R.~Dzhygadlo, A.~Gerhardt, M.~Krebs, D.~Lehmann, K.~Peters,
  G.~Schepers, C.~Schwarz, J.~Schwiening, M.~Traxler, M.~Düren L.~Schmitt~d,
  E.~Etzelmüller, K.~Föhl, A.~Hayrapetyan, K.~Kreutzfeld, J.~Rieke,
  M.~Schmidt, T.~Wasem, and C.~Sfienti.
\newblock Latest technological advances with {MCP-PMT}s.
\newblock \emph{Journal of Physics: Conference Series}, 2374:\penalty0 012128,
  2022.
\newblock \doi{10.1088/1742-6596/2374/1/012128}.
\newblock URL
  \url{https://iopscience.iop.org/article/10.1088/1742-6596/2374/1/012128/pdf}.

\bibitem[Daniel et~al.(2003)Daniel, Yahiaoui, Comtat, Jan, Kochebina, Martinez,
  Sergeyeva, Sharyy, Sung, and Yvon]{Daniel_2023}
Geoffrey Daniel, Mohamed~Bahi Yahiaoui, Claude Comtat, Sebastien Jan, Olga
  Kochebina, Jean-Marc Martinez, Viktoriya Sergeyeva, Viatcheslav Sharyy,
  Chi-Hsun Sung, and Dominique Yvon.
\newblock Deep learning reconstruction with uncertainty estimation for $\gamma$
  photon interaction in fast scintillator detectors.
\newblock \emph{Submitted to Engineering Applications of Artificial
  Intelligence.}, 2003.
\newblock \doi{10.48550/arXiv.2310.06572}.
\newblock URL \url{https://arxiv.org/abs/2310.06572}.

\bibitem[Kwon et~al.(2021)Kwon, Ota, Berg, Hashimoto, Nakajima, Ogawa,
  Tamagawa, Omura, Hasegawa, and Cherry]{Kwon2021}
S.I. Kwon, R.~Ota, E.~Berg, F.~Hashimoto, K.~Nakajima, I.~Ogawa, Y.~Tamagawa,
  T.~Omura, T.~Hasegawa, and S.~R. Cherry.
\newblock {Ultrafast timing enables reconstruction-free positron emission
  imaging}.
\newblock \emph{Nature Photonics}, 15:\penalty0 914–--918, 2021.
\newblock \doi{https://doi.org/10.1038/s41566-021-00871-2}.
\newblock URL \url{https://www.nature.com/articles/s41566-021-00871-2}.

\bibitem[Chen et~al.(2024)Chen, Ma, Huang, Hua, Jin, Jin, Qian, Ren, Si, Sun,
  Wu, Wang, Wang, Wang, Wang, Wu, and Zhang]{Chen2024}
Lingyue Chen, Lishuang Ma, Guorui Huang, Zhehao Hua, Muchun Jin, Zhen Jin, Sen
  Qian, Ling Ren, Shuguang Si, Jianning Sun, Qi~Wu, Xingchao Wang, Yifang Wang,
  Zhi Wang, Ning Wang, Kai Wu, and Haoda Zhang.
\newblock Coincidence time resolution of 50 ps fwhm using a pair of multi-anode
  mcp-pmts with cherenkov radiator window.
\newblock \emph{Nuclear Instruments and Methods in Physics Research Section A:
  Accelerators, Spectrometers, Detectors and Associated Equipment},
  1062:\penalty0 169173, 2024.
\newblock \doi{https://doi.org/10.1016/j.nima.2024.169173}.
\newblock URL
  \url{https://www.sciencedirect.com/science/article/abs/pii/S0168900224000998}.

\end{thebibliography}

%Within the same dark/Faraday box, we placed a back-to-back 511 keV  $^{22}$Na source at a distance $D1$ to the detector module instead of the laser.  We utilized the SiPM-LYSO gamma spectrometer shown above, placed at a distance $D2$ to the gamma source. 
%A diagram of such a setup is shown in Fig. \ref{fig:gammas_setup}. 
%Data acquisition is trigerred on.....
%The very small solid angle of the SiPM-LYSO spectrometer is a compromize between....
%We chose ..... 
%This "electronic collimation" allows to select 511 keV photons impinging spots sizes of $\sim$1.5 mm. 
%The SiPM-LYSO spectrometer and the gamma source move together along the to axis of the Clearmind prototype, thus....
%We thus select event happening.....
%The small solid angle of the SiPM induce low efficiency in data taking, thus each data point required a 60 minutes run.
\end{document}